\documentclass{jfm}

\usepackage{graphicx}
\usepackage{natbib}
\usepackage{color}

\ifCUPmtlplainloaded \else
  \checkfont{eurm10}
  \iffontfound
    \IfFileExists{upmath.sty}
      {\typeout{^^JFound AMS Euler Roman fonts on the system,
                   using the 'upmath' package.^^J}%
       \usepackage{upmath}}
      {\typeout{^^JFound AMS Euler Roman fonts on the system, but you
                   dont seem to have the}%
       \typeout{'upmath' package installed. JFM.cls can take advantage
                 of these fonts,^^Jif you use 'upmath' package.^^J}%
      }
  \else
  \fi
\fi

\ifCUPmtlplainloaded \else
  \checkfont{msam10}
  \iffontfound
    \IfFileExists{amssymb.sty}
      {\typeout{^^JFound AMS Symbol fonts on the system, using the
                'amssymb' package.^^J}%
       \usepackage{amssymb}%

      }{}
  \fi
\fi

\ifCUPmtlplainloaded \else
  \IfFileExists{amsbsy.sty}
    {\typeout{^^JFound the 'amsbsy' package on the system, using it.^^J}%
     \usepackage{amsbsy}}
    {\providecommand\boldsymbol[1]{\mbox{\boldmath $##1$}}}
\fi

\newsavebox{\astrutbox}
\sbox{\astrutbox}{\rule[-5pt]{0pt}{20pt}}

\newcommand{\drawline}[2]{\raisebox{2.5pt}{\vbox{\hrule width #1 pt height #2 pt}}}
\newcommand{\spacce}[1]{\hspace{#1pt}}
\newcommand{\solid}{\nobreak\mbox{\drawline{24}{0.5}\spacce{2}}}

\newcommand{\aline}{\drawline{3.429}{0.5}\spacce{3.429}}
\newcommand{\bline}{\drawline{3.429}{0.5}\spacce{2}}
\newcommand{\dashed}{\nobreak\mbox{\aline\aline\aline\bline}}
\newcommand{\ccline}{\drawline{5.5}{0.5}\spacce{2}\drawline{1}{0.5}\spacce{2}}
\newcommand{\dline}{\drawline{5.5}{0.5}\spacce{2}}
\newcommand{\dashdot}{\nobreak\mbox{\ccline\ccline\dline}}
\newcommand{\eline}{\drawline{0.857}{0.5}\spacce{1.714}}
\newcommand{\fline}{\drawline{0.857}{0.5}\spacce{2}}
\newcommand{\dotted}{\nobreak\mbox{\eline\eline\eline\eline\eline\eline\eline\eline\eline\fline}}

\newcommand{\gline}{\drawline{3.667}{0.5}\spacce{1.5}\drawline{1}{0.5}\spacce{1.5}\drawline{1}{0.5}\spacce{1.5}}
\newcommand{\hhhline}{\drawline{3.667}{0.5}\spacce{2}}
\newcommand{\dashdotdot}{\nobreak\mbox{\gline\gline\hhhline}}

\newcommand\calD{\calD} 

\def\calD    {\mathcal{D}}

\title[Self-sustaining process of minimal attached eddies in turbulent channel flow]
      {Self-sustaining process of minimal attached eddies in turbulent channel flow}

\author[Y. Hwang and Y. Bengana]%
{Yongyun Hwang$^1$\thanks{E-mail address for correspondence: y.hwang@imperial.ac.uk} \and Yacine Bengana$^2$}

\affiliation{$^1$Department of Aeronautics, Imperial College London, \\
South Kensington, London SW7 2AZ, UK\\[\affilskip]
$^2$D\'epartment de M\'ecanique, Universit\'e de Pierre et Marie Curie, \\
4 Place Jussieu, 75005 Paris, France}

\pubyear{}
\volume{}
\pagerange{}
\date{23 March 2016 ~ and in revised form ??}
\usepackage[hidelinks,breaklinks]{hyperref}
\begin{document}

\maketitle

\begin{abstract}
It has been recently shown that the energy-containing motions (i.e. coherent structures) in turbulent channel flow exist in the form of Townsend's attached eddies by a numerical experiment which simulates the energy-containing motions only at a prescribed spanwise length scale using their self-sustaining nature (Hwang, 2015, \emph{J. Fluid Mech.}, \textbf{767}, p254). In the present study, a detailed investigation of the self-sustaining process of the energy-containing motions at each spanwise length scale (i.e. the attached eddies) in the logarithmic and outer regions is carried out with an emphasis on its relevance to `bursting', which refers to an energetic temporal oscillation of the motions (Flores \& Jim\'enez, 2010, \emph{Phys. Fluids}, \textbf{22}, 071704). It is shown that the attached eddies in the logarithmic and outer regions, composed of streaks and quasi-streamwise vortical structures, bear the self-sustaining process remarkably similar to that in the near-wall region: i.e. the streaks are significantly amplified by the quasi-streamwise vortices via the lift-up effect; the amplified streaks subsequently undergo a `rapid streamwise meandering motion', reminiscent of streak instability or transient growth, which eventually results in breakdown of the streaks and regeneration of new quasi-streamwise vortices. For the attached eddies at a given spanwise length scale $\lambda_z$ between $\lambda_z^+\simeq 100$ and $\lambda_z\simeq 1.5h$, the single turn-over time period of the self-sustaining process is found to be $T u_\tau/\lambda_z\simeq 2$ ($u_\tau$ is the friction velocity), which corresponds well to the time scale of the bursting. Two additional numerical experiments, designed to artificially suppress the lift-up effect and the streak meandering motions, respectively, reveal that these processes are essential ingredients of the self-sustaining process of the attached eddies in the logarithmic and outer regions, consistent with several previous theoretical studies. It is also shown that the artificial suppression of the lift-up effect of the attached eddies in the logarithmic and outer regions leads to substantial amounts of turbulent skin-friction reduction.
\end{abstract}

\section{Introduction}

\cite{Townsend1961,Townsend1976} originally introduced the concept of the `attached eddy' to describe the energy-containing motions (i.e. coherent structures) populating the logarithmic region of wall-bounded turbulent flows. Given the length scale of the logarithmic region, the size of the energy-containing motions there would be proportional to the distance of their centre from the wall. This implies that, at least, some parts of the energy-containing motions would reach the wall, and, in this sense, they are expected to be `attached' to the wall. Under the assumption that these energy-containing motions in the logarithmic region (i.e. attached eddies) are statistically self-similar to one another, \cite{Townsend1976} theoretically predicted that turbulence intensity of the wall-parallel velocity components would exhibit a logarithmic wall-normal dependence as the mean-velocity profile does. The original theory of \cite{Townsend1976} has been significantly refined over a number of years by Perry and coworkers \cite[e.g.][among many others]{Perry1982,Perry1986,Perry1995,Nickels2005}, who have elaborated to develop a self-consistent model predicting the statistics of wall-bounded turbulent flows with a structural model of the attached eddy in the form of a hairpin or $\Lambda$ vortex. Especially, these authors further predicted and verified the emergence of $k_x^{-1}$ law in the spectra of wall-parallel velocity components ($k_x$ is the streamwise wavenumber), which would be linked to the logarithmic wall-normal dependence of turbulence intensity of these velocity components \cite[]{Perry1982}.

A growing body of evidence, which supports the theoretical predictions made with the attached eddy hypothesis, has emerged, especially for the last decade. The logarithmic growth of the near-wall streamwise turbulence intensity with the Reynolds number is an example of this, as it would be given by extending Townsend's prediction to the near-wall region \cite[]{Marusic2003}. The linear growth of the spanwise integral length scale with the distance from the wall has also been understood as another important evidence \cite[]{Tomkins2003,delAlamo2004,Monty2007}, as this indicates that the spanwise size of the energy-containing motions in the logarithmic region is proportional to the distance from the wall. Lastly, the logarithmic dependence of turbulence intensity of the wall-parallel velocity components has recently been indeed confirmed at sufficiently high Reynolds numbers \cite[]{Jimenez2008,Marusic2013,Orlandi2015}, suggesting that the energy-containing motions in wall-bounded turbulent flow would indeed be organised in the form of Townsend's attached eddies.

Very recently, the first author of the present study managed to compute the statistics of the attached eddies at a given length scale with a numerical experiment performed at moderate Reynolds numbers (${Re}_\tau \simeq 1000 \sim 2000$ where ${Re}_\tau$ is the friction Reynolds number) \cite[]{Hwang2015}. Built upon the linearly growing nature of the spanwise integral length scale with the distance from the wall, the numerical experiment is designed to isolate the energy-containing motions only at a given spanwise length scale using their `self-sustaining nature' \cite[]{Hwang2010c,Hwang2011}. The isolated self-sustaining energy-containing motions were found to be self-similar with respect to their spanwise length scale, and their statistical structures were remarkably similar to those of the attached eddies, demonstrating the existence of Townsend's attached eddies as energy-containing motions in wall-bounded turbulent flows. It was also shown that the single attached eddy is composed of two distinct elements, one of which is a long streaky motion and the other is a relatively compact streamwise vortical structure. At the considered Reynolds numbers, the size and the wall-normal location of the former streaky motion, the major carrier of the streamwise turbulence intensity, are roughly self-similar along
\begin{subequations}\label{eq:1.1}
\begin{equation}\label{eq:1.1a}
y\simeq 0.1\lambda_z~\mathrm{and}~\lambda_x \simeq 10 \lambda_z,
\end{equation}
where $y$ is the wall-normal direction, $\lambda_x$ the streamwise length scale, and $\lambda_z$ the spanwise length scale, while those of the latter vortical structure, carrying all the velocity components, also self-similarly scale with
\begin{equation}\label{eq:1.1b}
y\simeq 0.5\sim 0.7 \lambda_z~\mathrm{and}~\lambda_x \simeq 2\sim 3 \lambda_z.
\end{equation}
\end{subequations}
The scaling also reveals that the smallest attached eddy, given with $\lambda_z^+\simeq 100$, is a near-wall coherent motion in the form of a near-wall streak and several quasi-streamwise vortices aligned to it \cite[]{Hwang2013}, while the largest one, given with $\lambda_z\simeq 1.5h$ in a turbulence channel with the half height $h$, is an outer motion composed of a very-large-scale motion (VLSM) and several large-scale motions (LSMs) aligned to it \cite[]{Hwang2015}. The attached eddies at intermediate length scales, the size of which is proportional to their distance from the wall, are also in the form of a long streak and several quasi-streamwise vortices, and contribute to the logarithmic region by filling the gap caused by the separation between the inner and outer length scales. It should be mentioned that this scenario incorporates all the coherent structures currently known and exhibits the behaviour consistent with the spectra of all the velocities and Reynolds stress (including $k_x^{-1}$ spectra), providing an integrated description for the coherent structures within the attached eddy scenario \cite[for further details, see also][]{Hwang2015}.

The present study is an extension of the work by \cite{Hwang2015}, and, in particular, it is aimed to explore the `dynamics' of the self-sustaining attached eddies and its relevance to the corresponding motions under fully turbulent environment. For this purpose, in the present study, we consider the smallest computational box which allows for the self-sustaining mechanism of each of the attached eddies (i.e. the minimal unit). We then examine the detailed physical processes in the minimal unit by comparing them with those under fully-developed turbulence. Particular emphasis of the present study is given to relating the self-sustaining process of the attached eddies with the dynamical behaviour often referred to as `bursting' in the logarithmic and outer regions \cite[]{Flores2010}. To this end, we introduce two additional numerical experiments, designed to examine the physical processes composing the self-sustaining process of the attached eddies in the logarithmic and outer regions.

The paper is organised as follows. In $\S$\ref{sec:2}, we briefly introduce the numerical methods of extracting self-sustaining attached eddies at a given spanwise length scale \cite[see also][for further details]{Hwang2015}. A careful observation of the instantaneous flow field is then followed in $\S$\ref{sec:3} with examination of auto- and cross-correlation functions of several flow variables of interest. Especially in this section, we show that the bursting in the logarithmic and outer regions is likely a consequence of the self-sustaining process of the attached eddies. In $\S$\ref{sec:4}, a comprehensive discussion on the self-sustaining mechanism of the attached eddies is given with the two numerical experiments mentioned. The paper finally concludes in $\S$\ref{sec:5}.

\section{Numerical experiment}\label{sec:2}

\subsection{Computation of self-sustaining attached eddies}\label{sec:2.1}
We consider a turbulent channel, in which we denote $x$, $y$, and $z$ as the streamwise, wall-normal, and spanwise directions, respectively. The upper and lower walls are set to be at $y=0$ and $y=2h$, respectively, where $h$ is the half height of the channel. The numerical experiment in the present study is performed using the Navier-Stokes solver in our previous studies \cite[]{Hwang2010c,Hwang2011,Hwang2013,Hwang2015}. In this solver, the streamwise and spanwise directions are discretised using the Fourier-Galerkin method with 2/3 rule, and the wall-normal direction is discretised using second-order central difference. Time integration is performed semi-implicitly using the Crank-Nicolson method for the terms with wall-normal derivatives and a third-order low-storage Runge-Kutta method for the rest of the terms. The computations are performed by imposing a constant volume flux across the channel.

As mentioned, the present numerical experiment is largely based on the previous work by \cite{Hwang2015} (the reader is strongly recommended to refer to this paper in which the numerical technique introduced in this section is fully verified). In turbulent channel flow, the spanwise length scale exhibits a linear growth with the distance from the wall (i.e. $y$) in the range from $\lambda_z^+\simeq 100$ to $\lambda_z \simeq 1.5h$ \cite[]{delAlamo2004}. The spanwise spectra of the streamwise velocity clearly reveal this feature, and are approximately aligned along {$y\sim 0.1\lambda_z$} \cite[]{delAlamo2004,Hwang2015}. The key idea of \cite{Hwang2015} is to design a numerical simulation only for the energy-containing motions at a prescribed spanwise length $\lambda_z=\lambda_{z,0}$ between $\lambda_z^+\simeq 100$ and $\lambda_z \simeq 1.5h$ by artificially removing the motions at the other spanwise length scales (i.e. $\lambda_z\ne \lambda_{z,0}$)

First, the removal of the motions, the spanwise size of which is larger than the prescribed spanwise length $\lambda_{z,0}$ (i.e. $\lambda_z>\lambda_{z,0}$), is implemented by setting the size of the spanwise computational domain $L_z$ to be identical to the prescribed spanwise length scale $\lambda_{z,0}$ (i.e. $L_z= \lambda_{z,0}$), while keeping a sufficiently long streamwise domain $L_x$. However, even in such a computational domain, the motions at $\lambda_z>\lambda_{z,0}$ are not completely removed, as a spurious motion, uniform along the spanwise direction, would still survive \cite[]{Hwang2013}. This spurious motion is subsequently removed by taking the approach in \cite{Hwang2013} which explicitly filters out the spanwise uniform components by setting the right-hand side of the discretised momentum equation at each Runge-Kutta substep as follows:
\begin{eqnarray}\label{eq:2.1}
\widehat{\mathrm{RHS}}_{x} (y;k_x\neq0,k_z=0)=0, \nonumber\\
~\widehat{\mathrm{RHS}}_{y} (y;k_x\neq0,k_z=0)=0,
\end{eqnarray}
where the RHS$_i$ the right-hand side of  $i$-component of the discretized momentum equation, $~\widehat{\cdot}~$ denotes Fourier-transformed state in $x$ and $z$ directions, and $k_x$ and $k_z$ are the streamwise and spanwise wavenumbers, respectively. We note that this technique is only aimed at removing the two-dimensional spurious motion populating the $x$-$y$ plane, thus no action is taken for $\widehat{\mathrm{RHS}}_{z}$ \cite[for further details, see also][]{Hwang2013}.

Once the motions at $\lambda_z>\lambda_{z,0}$ are removed, the motions at $\lambda_z<\lambda_{z,0}$ are subsequently quenched using an over-damped large-eddy simulation (LES) \cite[]{Hwang2010c,Hwang2013,Hwang2015}. For the given scope of the present study, it is very important for the residual stress model of the LES not to generate any possible energy transfer to the resolved quantity. Therefore, as in \cite{Hwang2015}, we consider the static Smagorinsky model in which the residual stress is modelled in a purely diffusive form:
\begin{subequations}\label{eq:2.2}
\begin{equation}\label{eq:2.2a}
\tilde{\tau}_{ij}-\frac{\delta_{ij}}{3}\tilde{\tau}_{kk}=-2\nu_t \tilde{S}_{ij},
\end{equation}
with
\begin{equation}\label{eq:2.2b}
\nu_t =(C_s \tilde{\Delta})^2\tilde{\mathcal{S}}\mathcal{D},
\end{equation}
\end{subequations}
where $\tilde{\cdot}$ denotes the filtered quantity, $S_{ij}$ the strain rate tensor, $C_s$ the Smagorinsky constant, $\tilde{\Delta}=(\tilde{\Delta}_1\tilde{\Delta}_2\tilde{\Delta}_3)^{1/3}$ the nominal filter width, $\tilde{\mathcal{S}}=(2\tilde{S}_{ij}\tilde{S}_{ij})^{1/2}$ the norm of the strain rate tensor, and $\mathcal{D}=1-\exp[-(y^+/A^+)^3]$ is the van Driest damping function \cite[]{Hartel1998}. The removal of the motions at $\lambda_z<\lambda_{z,0}$ is subsequently carried out by increasing $C_s$ with the increment $\Delta C_s=0.05$ until we isolate the self-sustaining motions at $\lambda_z=\lambda_{z,0}$. We note that this approach is equivalent to increasing the filter width of LES without {losing} actual resolution \cite[]{Mason1986}. Finally, it should be mentioned that this approach itself does not strongly depend on the choice of the eddy viscosity model \cite[]{Hwang2010c,Hwang2011}.

The smallest computational domain, which allows for survival of the isolated self-sustaining attached eddies at a given spanwise length scale $\lambda_{z,0}$ (i.e. the minimal unit), is then sought. Since the spanwise size of the target attached eddies determines the spanwise computational domain $L_z$ (i.e. $L_z=\lambda_{z,0}$), only the size of the streamwise computational domain $L_x$ needs to be examined by gradually reducing it from a sufficiently large value. Such a search of the minimal unit of the self-sustaining attached eddies at a given spanwise length scale was previously performed in \cite{Hwang2010c} with the present over-damped LES, but without the filtering (\ref{eq:2.1}). In the present study, we therefore repeat this search again to check any possible effect of (\ref{eq:2.1}). It has been found that the filtering {(\ref{eq:2.1})} does not significantly affect the size of the minimal unit, resulting in the following streamwise computational domain size for the self-sustaining attached eddies at $\lambda_{z,0}=L_z$:
\begin{equation}\label{eq:2.3}
L_x =2L_z.
\end{equation}
In the present study, all the simulations are performed by keeping with this aspect ratio of the computational domain (see also table \ref{tab1}). {We note that the use of this minimal streamwise computational domain is not a great limitation of the present study, as the minimal domain is found not to significantly change the second-order statistics of the isolated self-sustaining attached eddies at $\lambda_{z,0}=L_z$. A detailed discussion on this issue is given in Appendix \ref{sec:appendixB} where the second-order statistics of the attached eddies with the minimal unit are compared with those with a long streamwise domain in \cite{Hwang2015}. Finally, it should be mentioned that the filtering of the uniform motion along the streamwise direction, such as (\ref{eq:2.1}) for the spanwise direction, is not implemented. As we shall see in $\S$\ref{sec:3} and $\S$\ref{sec:4}, the streamwise uniform Fourier mode in the minimal unit mainly resolves the long streaky motion observed in a long computational domain. Indeed, we will see that the streamwise uniform Fourier mode is strongly correlated with the streaky motion (see $\S$\ref{sec:3}), and inhibition of the related physical process destroys the self-sustaining process in the minimal unit (see $\S$\ref{sec:4}).}

\begin{table}
  \begin{center}
\def~{\hphantom{0}}
  \begin{tabular}{lccccccccc}
   Case    & $~Re_m~$   &  $~Re_\tau~$& $~L_x/h~$  & $~L_z/h~$  & 	$~N_x \times N_y \times N_z~$ & $~C_s~$ &  $T_{avg}u_{\tau}/h$  \\[3pt]

   \hline
   $L950a~$ & $~38133~$& $~941~$  & $~1.5~$ & $~0.75~$  &   $~24 \times 81 \times 24~$ & $~0.05~$  & $~247~$\\
   $SL950a~$ & $~38133~$&  $~936~$  & $~1.5~$ & $~0.75~$  &   $~24 \times 81 \times 24~$ & $~0.20~$ &  $~246~$\\ [3pt]

   $L950b~$ & $~38133~$& $~976~$  & $~2.0~$ & $~1.0~$  &  $~32 \times 81 \times 32~$ & $~0.05~$ & $~256~$\\
   $SL950b~$ & $~38133~$& $~1004~$  & $~2.0~$ & $~1.0~$  &  $~32 \times 81 \times 32~$ & $~0.25~$  &  $~263~$\\ [3pt]

   $O950~$  & $~38133~$& $~997~$ & $~3.0~$ & $~1.5~$   &   $~48 \times 81 \times 48~$ & $~0.05~$  & $~262~$\\
   $SO950~$  & $~38133~$& $~1152~$ & $~3.0~$ & $~1.5~$   &   $~48 \times 81 \times 48~$ & $~0.40~$  &  $~302~$ \\

   \hline
   $L1800a~$ & $~73333~$& $~1446~$    & $~0.75~$ & $~0.375~$   &  $~24 \times 129 \times 24~$ & $~0.05~$ & $~84~$\\
   $SL1800a~$  & $~73333~$& $~1438~$   & $~0.75~$ & $~0.375~$   &  $~24 \times 129 \times 24~$ & $~0.20~$ & $~84~$\\ [3pt]

   $L1800b~$ & $~73333~$& $~1606~$    & $~1.0~$ & $~0.5~$   &  $~32 \times 129 \times 32~$ & $~0.05~$ & $~93~$\\
   $SL1800b~$ & $~73333~$& $~1685~$    & $~1.0~$ & $~0.5~$   &  $~32 \times 129 \times 32~$ & $~0.30~$ &  $~98~$\\ [3pt]

   $L1800c~$ & $~73333~$& $~1745~$    & $~1.5~$ & $~0.75~$   &  $~48 \times 129 \times 48~$ & $~0.05~$ & $~102~$\\
   $SL1800c~$ & $~73333~$& $~1954~$    & $~1.5~$ & $~0.75~$   &  $~48 \times 129 \times 48~$ & $~0.40~$ &  $~114~$\\
  \end{tabular}
  \caption{Simulation parameters in the present study. Here, $Re_m$ is the Reynolds number based on the bulk velocity and $T_{avg}$ is the time interval for average. In the names of the simulations, $L$ indicates `logarithmic', $O$ `outer' and $S$ `self-sustaining'.}
  \label{tab1}
  \end{center}
\end{table}

The simulation parameters of the present study are summarised in table \ref{tab1}, where six different simulations are considered at two different Reynolds numbers ($Re_\tau\simeq 950$ and $Re_\tau\simeq 1800$). For each computational domain, we consider a pair of simulations for the purpose of comparison. One is with $C_s=0.05$, the value known to provide the best statistics compared to that of DNS \cite[]{Hartel1998} and is performed to track the attached eddies with background turbulence, and the other is with an increased $C_s$ and is to study the attached eddies surviving only through the self-sustaining mechanism. Typical spatial grid spacing of the simulations is set to be in the range of $\Delta x^+=50\sim 70$, $\Delta {y^+}_{\mathrm{min}}=1 \sim 2$, and $\Delta z^+=20 \sim 30$, in order to properly resolve the near-wall motions. We note that this grid spacing is finer than that in \cite{Hwang2015} where the numerical technique applied here is fully verified with very long $L_x$. Finally, it should be mentioned that significantly long average time ($T_{avg}$) are considered to ensure sufficient convergence of the time correlation functions introduced in $\S$\ref{sec:3}. The convergence of the correlation functions have also been checked by doubling $T_{avg}$, but it is not significantly improved.

\subsection{Validation and preliminary test}\label{sec:2.2}

\begin{figure} \vspace*{2mm}
\centering
\includegraphics[width=0.98\textwidth]{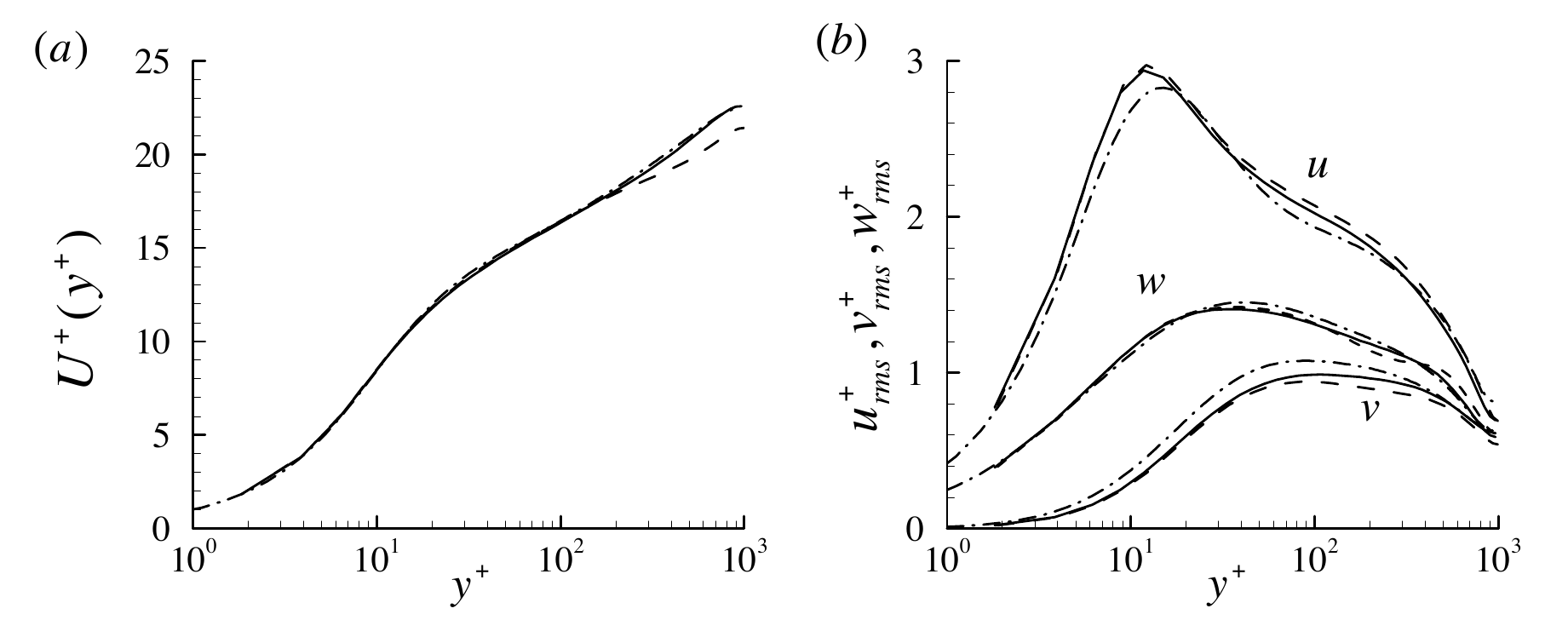}
\caption{$(a)$ Mean-velocity profile and $(b)$ turbulent velocity fluctuations of the reference simulation: \protect \solid, $O950$ without the filtering {(\ref{eq:2.1})}; \protect \dashed, $O950$; \protect \dashdot, DNS at $Re_{\tau}=934$ \cite[]{delAlamo2004}.}\label{fig1}
\end{figure}

To verify the present LES, a reference simulation for $C_s=0.05$ without applying (\ref{eq:2.1}) is first performed with the domain of $O950$: i.e. the minimal unit for the outer attached eddies \cite[]{Hwang2010c}. Figure \ref{fig1} shows the mean-velocity profile and turbulent velocity fluctuations of the reference simulation. The statistics of the simulation show reasonable agreement with those of DNS by \cite{delAlamo2004}, consistent with \cite{Lozano-Duran2014} who reported the effect of the computational domain size on the statistics. As discussed in $\S$\ref{sec:2.1}, the present numerical experiment also implements the filtering action (\ref{eq:2.1}) unlike the ordinary minimal-box simulation which only considers a spatially confined computation box. Therefore, the statistics of the simulation for $C_s=0.05$ with (\ref{eq:2.1}) is also checked (the dashed line in figure \ref{fig1}). Although the filtering (\ref{eq:2.1}) a little distorts the mean-velocity profile and the velocity fluctuations roughly for $y^+>150$ ($y>0.15h$), it does not yield significant change overall in the statistics. Further inspection is performed by computing the time correlation functions of interest defined in $\S$\ref{sec:3}, and reveals that the filtering (\ref{eq:2.1}) does not significantly change the dynamical process in the considered computational domain. For further details, the reader refers to Appendix \ref{sec:appendix} where a detailed comparison of the computed time correlation functions is made.

\begin{figure} \vspace*{2mm}
\centering
\includegraphics[width=0.98\textwidth]{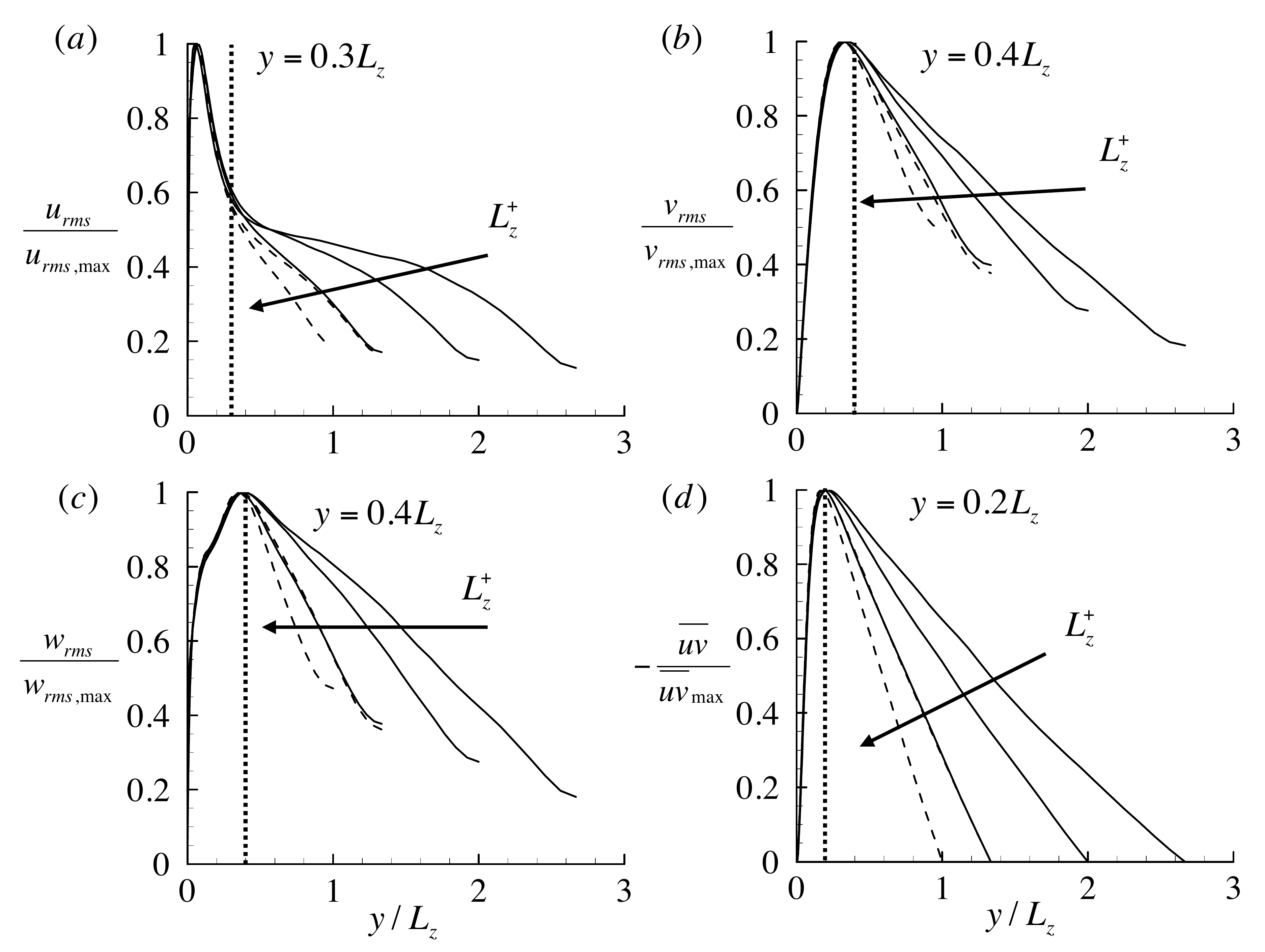}
\caption{Normalised second-order statistics of the attached eddies, scaled with the given spanwise length $L_z$: $(a)$ streamwise velocity; $(b)$ wall-normal velocity; $(c)$ spanwise velocity; $(d)$ Reynolds stress. Here, \protect \dashed, from $Re_{\tau} \simeq 950$ ($SL950a$, $SL950b$); \protect \solid, from $Re_{\tau} \simeq 1800$ ($SL1800a$, $SL1800b$, $SL1800c$).}\label{fig2}
\end{figure}

The only difference between the present numerical experiment and that in \cite{Hwang2015} is the streamwise size of the computational domain. To ensure that the change of the streamwise computational domain does not affect the self-similar nature of the computed attached eddies, the statistics of the attached eddies, obtained by implementing the technique in $\S$\ref{sec:2.1}, are also reported. Figure \ref{fig2} shows the normalised second-order statistics of the simulations in which the attached eddies at the given spanwise length scale are isolated (i.e. the simulations tagged with `$S$' in table \ref{tab1}). The computed statistics of the attached eddies are approximately self-similar with respect to their spanwise length scale $L_z$ below certain wall-normal locations: the streamwise velocity fluctuation is self-similar roughly below $y\simeq0.3L_z$ (figures \ref{fig2}$a$), the wall-normal and the spanwise velocity fluctuations are below $y\simeq0.4L_z$ (figures \ref{fig2}$b$ and $c$), and the Reynolds stress is below $y\simeq0.2L_z$ (figures \ref{fig2}$d$). Above these wall-normal locations, the second-order statistics depend on the box size, as in \cite{Hwang2015}. However, we note that the non-self-similar part of the fluctuations above these wall-normal locations is merely a fluctuation induced by the self-similar part in the wall-normal locations supposed to be `empty', due to the absence of the motions larger than the attached eddies of interest \cite[for further disucssion, see also][]{Hwang2013,Hwang2015}. Finally, it should be mentioned that the second-order statistics of the wall-parallel velocity components show non-negligibly large contributions to the region close to the wall (figures \ref{fig2}$a$, $c$), whereas those of the wall-normal velocity and the Reynolds stress are fairly small (figures \ref{fig2}$b$, $d$): {for example, at $y=0.05L_z$, the normalised streamwise and the spanwise velocity fluctuations are respectively given around 0.98 and 0.68 times of their maxima, whereas the wall-normal velocity and Reynolds stress are respectively found around 0.28 and 0.45 times of the maxima.} This is essentially due to the impermeability condition caused by the presence of the wall, and is exactly as in the statistical structure of single attached eddy hypothesised by \cite{Townsend1976}.

\section{Results}\label{sec:3}
\subsection{The largest attached eddies: large-scale and very-large-scale motions}\label{sec:3.1}
We first explore the temporal evolution of the largest attached eddies (i.e. the energy-containing motions at $\lambda_z=1.5h$) in the minimal unit ($O950$ and $SO950$ in table \ref{tab1}). As discussed in \cite{Hwang2015}, the largest attached eddy is composed of a VLSM (outer streaky motion) and LSMs (outer streamwise vortical structures) aligned with it. {In a large computational domain, the VLSM is typically featured to be a long structure mainly carrying a large amount of the streamwise turbulent kinetic energy, and this is quite similar to the behaviour of the streaks in the near-wall region \cite[]{Hwang2015}}. Therefore, in the {outer} minimal unit, in which only single VLSM would be resolved mainly by zero streamwise wavenumber due to its long length (i.e. $k_x=0$; see also figure \ref{fig6}), its temporal evolution would be well tracked by
\begin{subequations}\label{eq:3.1}
\begin{equation}\label{eq:3.1a}
E_{u}=\frac{1}{2V_h}\int_{\Omega_h} {u}^2~dV,
\end{equation}
where $u$ is the streamwise velocity fluctuation, $\Omega_h$ the lower (or upper) half of the computational domain of interest, and $V_h$ the volume of the half of the computational domain.

{On the contrary, the LSM is relatively short structure given with its streamwise length scale at $\lambda_x\simeq 3\sim 4h$. The structure at this length is identical to that often postulated as a hairpin vortex packet by Adrian and coworkers \cite[]{Adrian2007}. This structure is relatively isotropic, compared to the VLSM which mostly carries the streamwise velocity fluctuation, in the sense that it contains all the velocity components \cite[]{Hwang2015}. It is important to mention that this structure was found to share a number of statistical similarities with quasi-streamwise vortex in the near-wall region \cite[]{Hwang2015}. This suggests that the LSMs would likely to be quasi-streamwise vortical structures in the outer region, although they are much more disorganised due to small-scale turbulence associated with cascade and turbulent dissipation \cite[]{Jimenez2012}. It should be mentioned that this view is consistent with \cite{delAlamo2006b} who called this structure `tall-attached vortex cluster'.} Since the VLSM carries very little wall-normal or spanwise turbulent kinetic energy, in the minimal unit, the LSMs would be well characterised by either
\begin{equation}\label{eq:3.1b}
E_{v}=\frac{1}{2V_h}\int_{\Omega_h} {v}^2~dV,
\end{equation}
or
\begin{equation}\label{eq:3.1c}
E_{w}=\frac{1}{2V_h}\int_{\Omega_h} {w}^2~dV,
\end{equation}
\end{subequations}
where $v$ and $w$ are the wall-normal and spanwise velocity fluctuations, respectively. Later, we shall see that these two variables are indeed strongly correlated to each other (figures \ref{fig5}$c$ and $d$). {Here, it should also be pointed out that the use of the primitive variables in characterising the quasi-streamwise vortical structure instead of enstrophy or streamwise vorticity is intentionally introduced to track their behaviour more precisely. Given the significant amount of small-scale turbulence associated with energy cascade and turbulent dissipation in the logarithmic and the outer regions, the variables such as enstrophy or streamwise vorticity would not faithfully represent the large-scale organised vortical structure. Indeed, \cite{Jimenez2012} recently showed that such variables carry dominant energy around turbulent dissipation length scales.}

\begin{figure} \vspace*{2mm}
\centering
\includegraphics[width=0.98\textwidth]{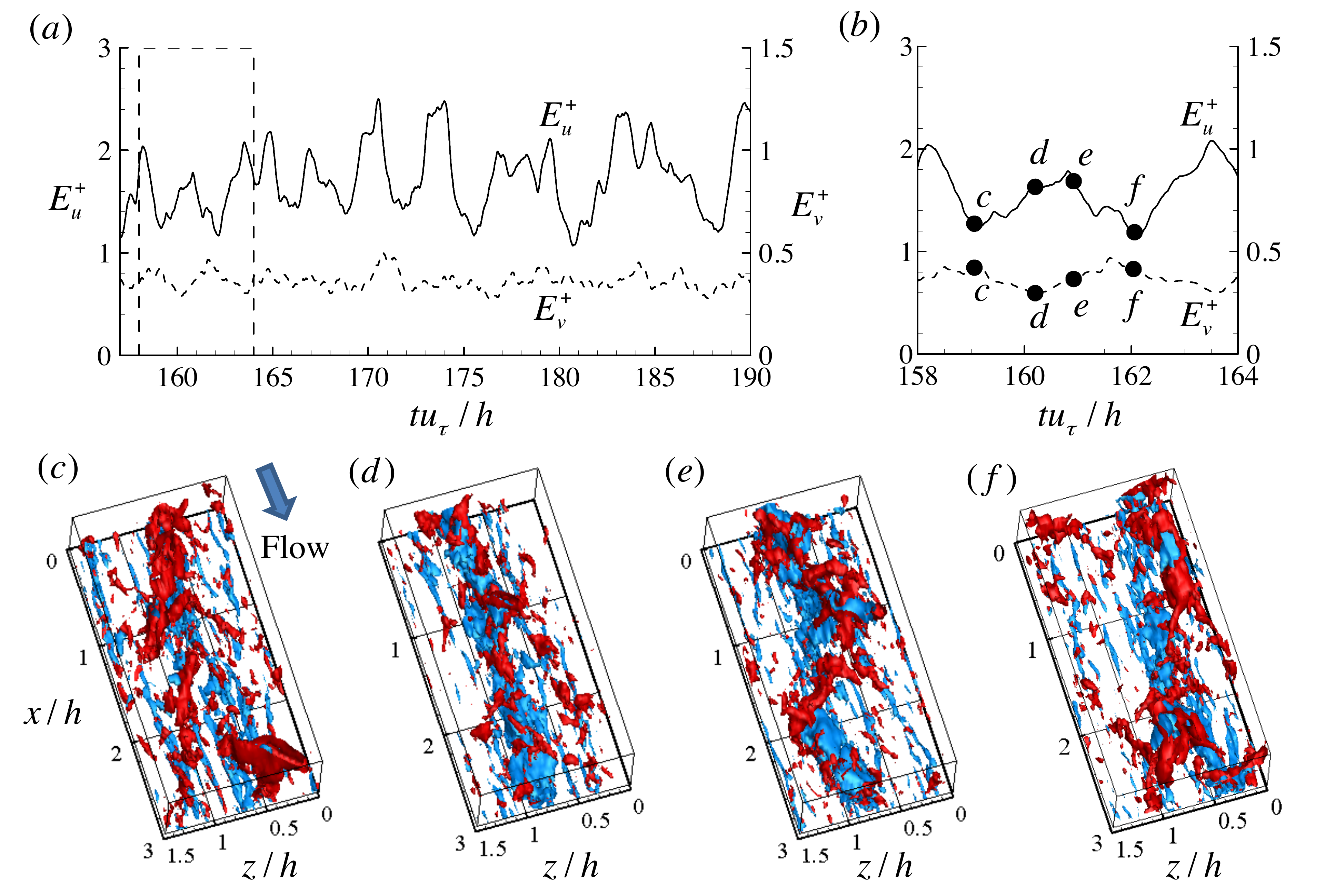}
\caption{Time evolution of the flow field of $O950$: (a) time trace of $E_{u}^+$ (solid) and $E_{v}^+$ (dashed); (b) magnification of (a) for $tu_{\tau}/h \in [158,164]$; $(c-f)$ the corresponding flow visualisation at $tu_{\tau}/h=159,160.2,160.9,162$. In $(c-f)$, the red and blue iso surfaces indicate ${u}^+=-3.2$ and $v^+=1.4$, respectively.}\label{fig3}
\end{figure}

A set of the flow fields of $O950$, in which the largest attached eddies are contained with other smaller-scale turbulent motions, are first examined. Figure \ref{fig3} shows the time trace of $E_u$ and $E_v$, and the corresponding evolution of the flow field in time. Both $E_u$ and $E_v$ show large-scale temporal oscillations with a time scale roughly at $Tu_\tau/h \simeq 2\sim 5$ (figure \ref{fig3}$a$). This is exactly the feature known as `bursting' by \cite{Flores2010}. A careful observation reveals that $E_u$ and $E_v$ oscillate with a certain phase difference: for example, at $tu_{\tau}/h=159$, $E_v$ is around the local maximum while $E_u$ being at fairly low-energy state (figure \ref{fig3}$b$), and, at $tu_{\tau}/h=160.2$, $E_v$ reaches the local minimum whereas $E_u$ becomes considerably large (figure \ref{fig3}$b$). Visualisation of the corresponding flow fields in figures \ref{fig3} $(c$-$f)$ suggests that this is likely due to the interactive dynamics between the VLSM (streak) and the LSMs (streamwise vortical structures). At $tu_{\tau}/h=159$ (figure \ref{fig3}$c$), the flow field exhibits a few fairly strong large-scale $v$ structures, which would be a part composing the LSMs in the minimal unit. On the other hand, $u$ structures in the flow field are fairly weak at this time, and they appear to be collectively gathered around the $v$ structures. At $tu_{\tau}/h=160.2$ (figure \ref{fig3}$d$), the $v$ structures become considerably weaken whereas the $u$ structures are significantly amplified, forming a strong streaky structure extending over the entire streamwise domain. This amplification of the streaky structure appears to be a consequence of the ejection of the streamwise momentum by the strong $v$ structures observed at $tu_{\tau}/h=159$ (figure \ref{fig3}$c$), clearly reminiscent of the `lift-up' effect predicted by previous theoretical studies \cite[]{delAlamo2006,Pujals2009,Cossu2009,Hwang2010b}. The amplified streaky structure subsequently meanders along the streamwise direction (figure \ref{fig3}$e$). This process eventually leads to regeneration of new $v$ structures with the breakdown of the streaky motion (figure \ref{fig3}$f$).

\begin{figure} \vspace*{2mm}
\centering
\includegraphics[width=0.98\textwidth]{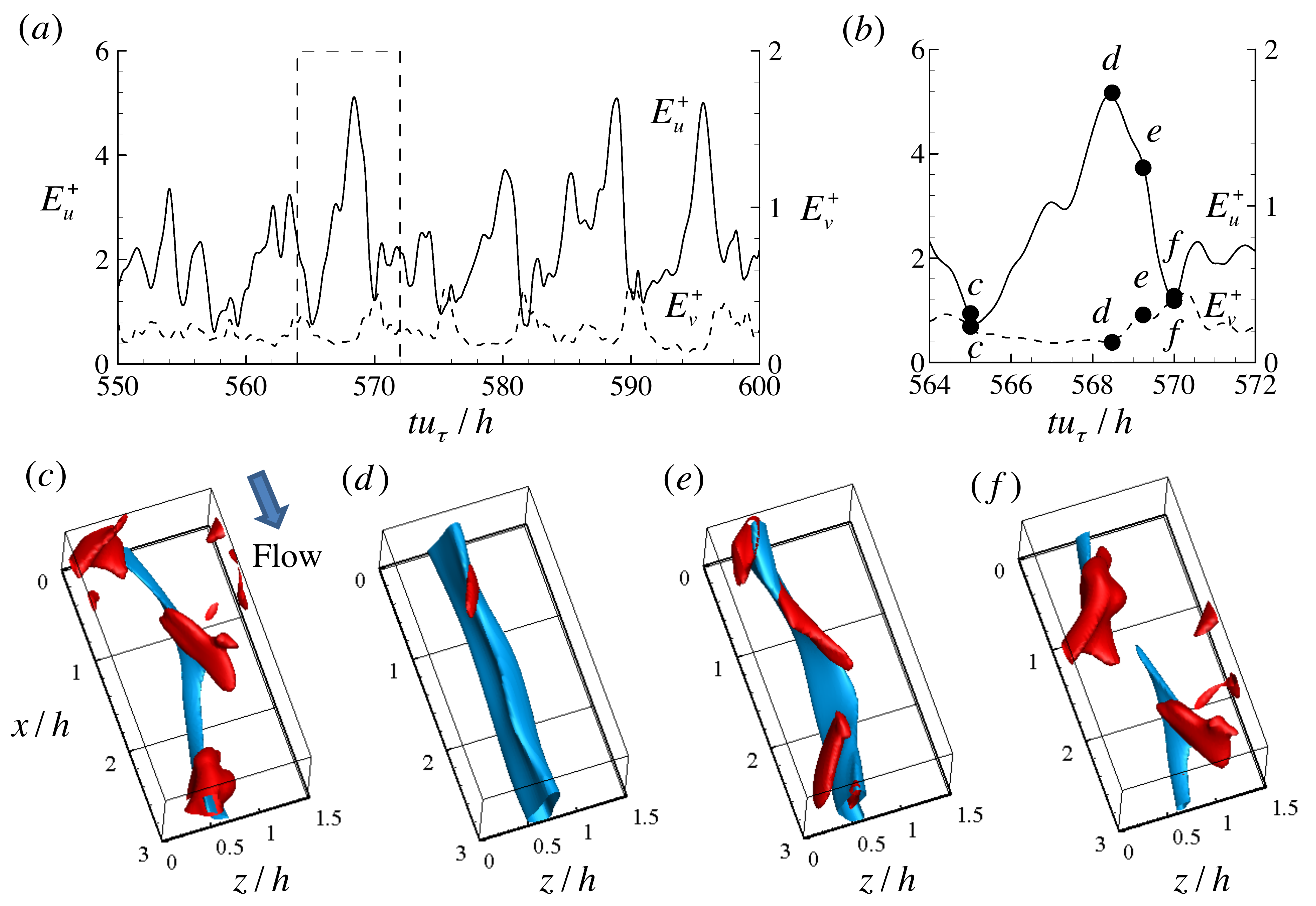}
\caption{Time evolution of the flow field of $SO950$: (a) time trace of $E_{u}^+$ (solid) and $E_{v}$ (dashed); (b) magnification of (a) for $tu_{\tau}/h \in [564,572]$; $(c-f)$ the corresponding flow visualisation at $tu_{\tau}/h=565,568.5,569.3,570$. In $(c-f)$, the red and blue iso surfaces indicate ${u}^+=-4$ and $v^+=1.5$, respectively.}\label{fig4}
\end{figure}

Qualitatively the same behaviour is observed in the over-damped simulation $SO950$, in which only the energy-containing motions at $\lambda_z=1.5h$ are isolated by replacing all the smaller scale motions with a crude eddy viscosity given with the increased $C_s$. Figure \ref{fig4} shows the time trace of $E_u$ and $E_v$, and the corresponding evolution of the flow field of $SO950$. As in $O950$ (figure \ref{fig3}$a$), both $E_u$ and $E_v$ exhibit temporal oscillations with the time scale at $Tu_{\tau}/h=2\sim 5$ (figure \ref{fig4}$a$). In this case, more prominent phase difference is observed between the oscillations of $E_u$ and $E_v$ (figures \ref{fig4}$a$ and $b$), presumably because the smaller-scale surrounding turbulent motions are removed in $SO950$. However, this could also be partially due to the increased oscillation amplitude of $E_u$ in $SO950$, which has previously been shown to be an artifact caused either by the crude eddy viscosity itself or by the lack of the energy-containing motions in the near-wall and the logarithmic regions \cite[]{Hwang2015}. Not surprisingly, the flow fields of $SO950$ are composed of much smoother structures than those of $O950$ (figures \ref{fig4}$c$-$f$). However, the temporal evolution of the $u$ and $v$ structures of $SO950$ is remarkably similar to that of $O950$: the $v$ structures significantly amplify the streaky $u$ structure via the lift-up effect (figures \ref{fig4}$c$ and $d$); the amplified streaky $u$ motion subsequently meanders along the streamwise direction (figure \ref{fig4}$e$); the $v$ structures are finally regenerated with breakdown of the streaky $u$ motion (figure \ref{fig4}$f$).

To statistically quantify this cyclic dynamical process, auto- and cross-correlations of several variables of interest are computed. It is also useful to introduce two additional variables, such that:
\begin{subequations}\label{eq:3.2}
\begin{equation}\label{eq:3.2a}
E_0=\int_{\Omega_{y,h}}|\hat{u}(y;k_x,k_z)|^2+|\hat{v}(y;k_x,k_z)|^2+|\hat{w}(y;k_x,k_z)|^2~dy,
\end{equation}
for $k_x=0$ and $k_z=2\pi/L_z$, and
\begin{equation}\label{eq:3.2b}
E_1=\int_{\Omega_{y,h}}|\hat{u}(y;k_x,k_z)|^2+|\hat{v}(y;k_x,k_z)|^2+|\hat{w}(y;k_x,k_z)|^2~dy,
\end{equation}
\end{subequations}
for $k_x=2\pi/L_x$ and $k_z=2\pi/L_z$. Here, $\hat{\cdot}$ denotes the Fourier-transformed state in the $x$ and $z$ directions, and $\Omega_{y,h}$ is the lower (or upper) half of the wall-normal domain. We note that, in (\ref{eq:3.2}), $E_0$ is introduced to compute the energy of the streamwise uniform component of the motions at $\lambda_z=1.5h$, while $E_1$ is to compute the energy of the first streamwise Fourier component, which would measure the meandering motion observed in figures \ref{fig3} and \ref{fig4}. The correlation functions are defined as
\begin{equation}
C_{ij}(\tau)=\frac{\langle E_i(t+\tau) E_j(t)\rangle}{\sqrt{\langle {E_i}^2(t) \rangle} \sqrt{\langle {E_j}^2(t) \rangle}},
\end{equation}
where $i,j=u,v,w,0,1$ and $\langle \cdot \rangle$ indicates average in time. In the present study, all the correlation functions are computed by averaging the lower and upper half channels.

\begin{figure} \vspace*{2mm}
\centering
\includegraphics[width=0.88\textwidth]{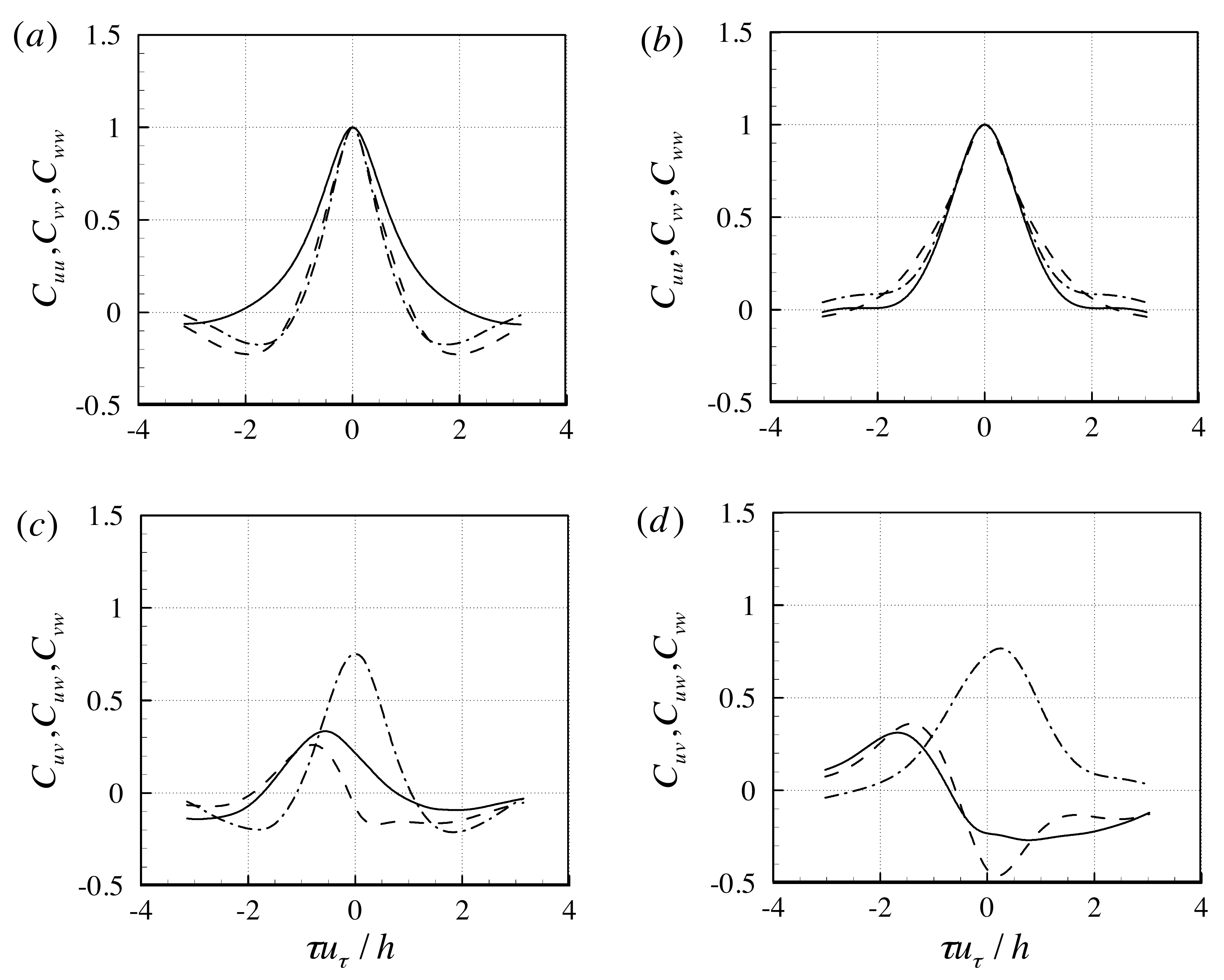}
\caption{Time correlation functions of $(a,c)$ $O950$ and $(b,d)$ $SO950$. In $(a,b)$, \protect \solid, $C_{uu}(\tau)$;  \protect \dashed, $C_{vv}(\tau)$; \protect \dashdot, $C_{ww}(\tau)$. In $(c,d)$, \protect \solid, $C_{uv}(\tau)$; \protect \dashed, $C_{uw}(\tau)$; \protect \dashdot, $C_{vw}(\tau)$.}\label{fig5}
\end{figure}

Figure \ref{fig5} shows the correlation functions computed with $E_u$, $E_v$, and $E_w$ from $O950$ and $SO950$. In the case of $O950$, all the auto-correlations decay to zero at $|\tau u_{\tau}/h|\simeq 1 \sim 2$, although $C_{uu}$ tends to drop a little more slowly than $C_{vv}$ and $C_{ww}$ (figure \ref{fig5}$a$). The size of the time interval, in which the auto-correlations remain positive, is $T u_{\tau}/h \simeq 2\sim 4$, and this roughly corresponds to the single period of the temporal oscillation of $E_u$ and $E_v$ (figure \ref{fig3}$a$). The cross-correlation $C_{uv}$ of $O950$ also reveals that there is indeed a phase difference between $E_u$ and $E_v$  (figure \ref{fig5}$c$): the peak of $C_{uv}$ is located at $\tau u_{\tau}/h \simeq -0.6$, indicating that $E_u$ statistically reaches its local extremum roughly $\Delta \tau u_{\tau}/h \simeq 0.6$ before $E_v$ reaches its local extremum. On the other hand, $E_v$ and $E_w$ are found to be strongly correlated to each other: $C_{vw}(\tau=0)\simeq 0.7$ and $C_{uw}$ also exhibits a peak at $\tau u_{\tau}/h \simeq -0.6$ as $C_{uv}$ does (figure \ref{fig5}$c$).

Qualitatively the same behaviour is observed in the correlation functions of $SO950$ in which only the energy-containing motions at $\lambda_z\simeq 1.5h$ are isolated (figures \ref{fig5}$b$ and $d$). All the auto-correlations drop to zero near $|\tau u_{\tau}/h|\simeq 1.5 \sim 2$ (figure \ref{fig5}$b$), resulting in the time interval of positive auto-correlation to be $T u_{\tau}/h \simeq 4$. As for $O950$, this time scale reasonably well quantifies single period of the temporal oscillations of $E_u$ and $E_v$ (figure \ref{fig4}$a$). The phase difference between $E_u$ and $E_v$ is also seen in $C_{uv}$ of $SO950$ (figure \ref{fig5}$d$) which shows the peak at $\tau u_{\tau}/h \simeq -1.7$. Finally, $E_v$ and $E_w$ are also found to be strongly correlated to each other, as seen in $C_{vw}$ (figure \ref{fig5}$d$).

\begin{figure} \vspace*{2mm}
\centering
\includegraphics[width=0.88\textwidth]{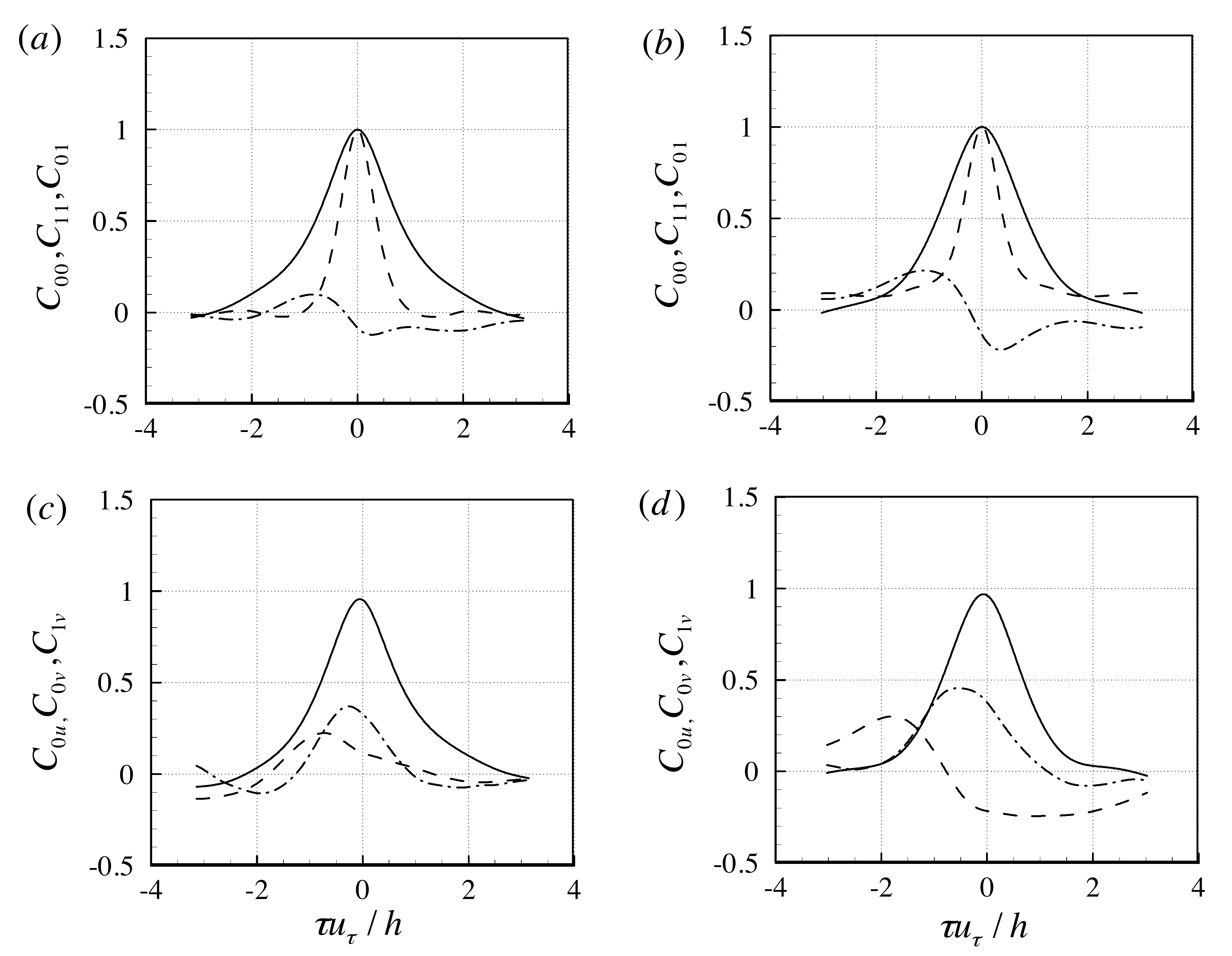}
\caption{Time correlation functions of $(a,c)$ $O950$ and $(b,d)$ $SO950$. In $(a,b)$, \protect \solid, $C_{00}(\tau)$;  \protect \dashed, $C_{11}(\tau)$;  \protect \dashdot, $C_{01}(\tau)$. In $(c,d)$, \protect \solid, $C_{0u}(\tau)$; \protect \dashed, $C_{0v}(\tau)$; \protect \dashdot, $C_{1v}(\tau)$.}\label{fig6}
\end{figure}

The correlation functions with $E_0$ and $E_1$ are also presented in figure \ref{fig6}. The auto- and cross-correlations obtained only with $E_0$ and $E_1$ (i.e. $C_{00}$, $C_{11}$ and $C_{01}$) are given in figures \ref{fig6} ($a$) and ($b$). The correlation functions of $O950$ show fairly similar behaviour to those of $SO950$: $C_{00}$ and $C_{11}$ of $O950$ exhibit similar time scales to those of $SO950$, and $C_{01}$ of both $O950$ and $SO950$ has a phase difference between $E_0$ and $E_1$ with $\Delta \tau  u_{\tau}/h =1$. {The phase difference between $E_0$ and $E_1$ implies that the streamwise uniform motion occurs first and the meandering motion follows, as in the instantaneous fields shown in figures \ref{fig3} and \ref{fig4}.} Here, it is interesting to note that the time scale of $C_{11}$ ($T u_{\tau}/h \simeq 2$) appears to be much smaller than that of $C_{00}$ as well as that of $C_{uu}$, $C_{vv}$ and $C_{ww}$ ($T u_{\tau}/h \simeq 4$; see also figures \ref{fig5}$a$ and $b$). This suggests that the meandering motion of the streak (figures \ref{fig5}$c$ and \ref{fig6}$c$), which would be characterised by $E_1$, is a rapid process which persists only for roughly half of the bursting period.

Cross-correlations are further computed by also considering $E_u$, $E_v$, and $E_w$, as shown in figures \ref{fig6} ($c$) and ($d$). The correlation $C_{0u}$ shows that $E_0$ and $E_u$ are strongly correlated to each other in both $O950$ and $SO950$, indicating that the streamwise uniform mode ($k_x=0$) at $\lambda_z=1.5h$ well represents the streaky motion (i.e. VLSM) in the minimal unit. This feature can also be confirmed from $C_{0v}$ of $O950$ and $SO950$, which behaves very similarly to $C_{uv}$ (figures \ref{fig5}$c$ and $d$). Finally, the peak of $C_{1v}$ for both $O950$ and $SO950$ is found roughly at $\tau u_{\tau}/h \simeq -1$. This indicates that the meandering motion of the streak statistically appears before the vortical structures (LSMs) are fully amplified, consistent with the instantaneous flow fields given in figures \ref{fig3} and \ref{fig4}.

Despite many qualitative similarities between the dynamical behaviours of the energy-containing motions in $O950$ and $SO950$, it should be pointed out that the correlation functions of the two simulations ($O950$ and $SO950$) are not quantitatively the same with each other, although this is not very surprising given the aggressive nature of the present numerical experiment: for example, the time intervals of $C_{vv}>0$ and $C_{ww}>0$ of $O950$ are smaller than those of $SO950$ (figures \ref{fig5}$a$ and $b$), and the peak locations of $C_{uv}$, $C_{uw}$, and $C_{0v}$ of $O950$ are a little different from those of $SO950$ (figures \ref{fig5}$c$, \ref{fig5}$d$, \ref{fig6}$c$ and \ref{fig6}$d$). Apparently, these differences would stem from two possible origins: one is that $E_u$, $E_v$ and $E_w$ of $O950$ contain the effect of the surrounding smaller-scale turbulent motions, and the other is that the dynamical behaviour of the motions in $SO950$ is probably a little distorted by the artificially increased eddy viscosity of $SO950$. The issue of which of the two more dominantly yields the differences could be clarified by further inspecting $C_{00}$, $C_{11}$, and $C_{01}$ (figures \ref{fig6}$a$ and $b$), as they are, in a way, obtained by applying cut-off filters to both of $O950$ and $SO950$. It appears that the correlation functions of $O950$ are not exactly the same as those of $SO950$, suggesting that increasing $C_s$ a little distorts the motions at $\lambda_z=1.5h$. However, the times scales of these correlation functions for $O950$ do not appear to be significantly different from those for $SO950$. This implies that the correlation functions obtained with $E_u$, $E_v$ and $E_w$ of $O950$ do not precisely represent the dynamical behaviour of the largest attached eddies at $\lambda_z=1.5h$ due to the contribution by the surrounding smaller-scale turbulent fluctuations.

\subsection{The attached eddies in the logarithmic region}\label{sec:3.2}

\begin{figure} \vspace*{2mm}
\centering
\includegraphics[width=0.88\textwidth]{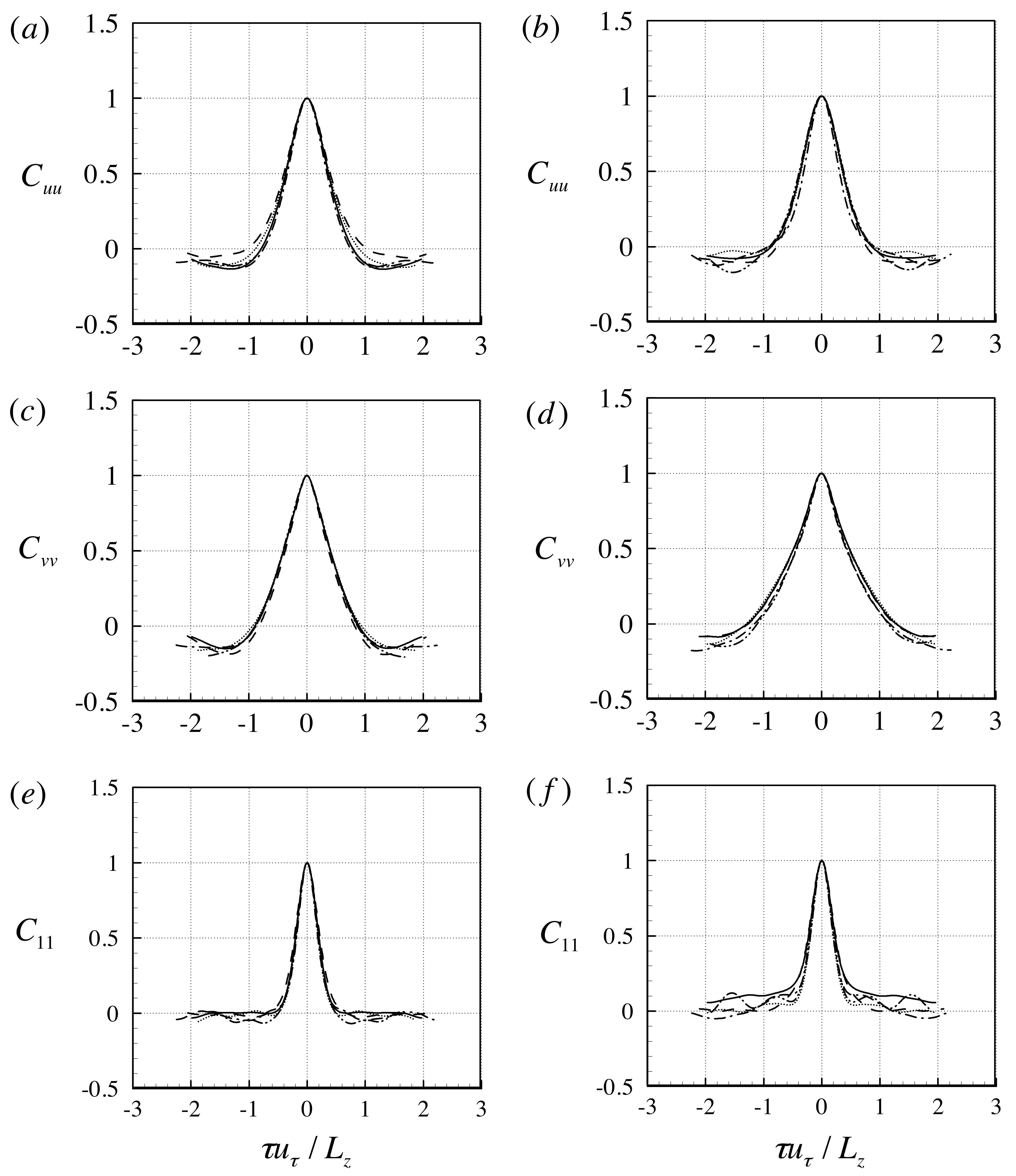}
\caption{Auto-correlation functions: $(a,b)$ $C_{uu}(\tau)$; $(c,d)$ $C_{vv}(\tau)$; $(e,f)$ $C_{11}(\tau)$. In $(a,c,e)$, \protect \solid $L950a$; \protect \dashed $L950b$; \protect \dashdot $L1800a$; \protect \dashdotdot $L1800b$; \protect \dotted $L1800c$, while, in $(b,d,f)$, \protect \solid $LS950a$; \protect \dashed $LS950b$; \protect \dashdot $LS1800a$; \protect \dashdotdot $LS1800b$; \protect \dotted $LS1800c$.}\label{fig7}
\end{figure}

Now, we consider the attached eddies in the logarithmic region. As shown in $\S$\ref{sec:2.2}, the attached eddies in the logarithmic region generate self-similar statistical structure with respect to the spanwise length scale $\lambda_{z,0}(=L_z)$. This self-similar part, which contains the essential physical process of the energy-containing motions in the logarithmic region, appears mainly below $y\simeq 0.3\sim 0.4L_z$ (figure \ref{fig2}). To trace the dynamical behaviour of this self-similar part more precisely, the definitions of $E_u$, $E_v$, $E_w$, $E_0$, and $E_1$ given in (\ref{eq:3.1}) and (\ref{eq:3.2}) are a little modified, so that the wall-normal domain of the integration becomes only $[0,2/3L_z]$. We note that this new definition is still consistent with (\ref{eq:3.1}), as the wall-normal domain of the integration becomes $[0,1]$ for $L_z=1.5h$, which is the spanwise length scale of the VLSM and the LSM.

Inspection of the instantaneous flow fields of all the simulations on the logarithmic region (see also table \ref{tab1}) reveals that basically the same dynamical process occurs in the attached eddies, given with streaks and quasi-streamwise vortical structures, at each of the spanwise length scales belonging to the logarithmic region: i.e. amplification of the streaks by the quasi-streamwise vortical structures, subsequent meandering of the streaks along the streamwise direction, and breakdown of the streaks with regeneration of the quasi-streamwise vortical structures. To avoid repetition of the same discussion given in $\S$\ref{sec:3.1}, here we only report auto- and cross-correlation functions of the simulations concerning the logarithmic region with focus on scaling of the computed correlation functions.

\begin{figure} \vspace*{2mm}
\centering
\includegraphics[width=0.88\textwidth]{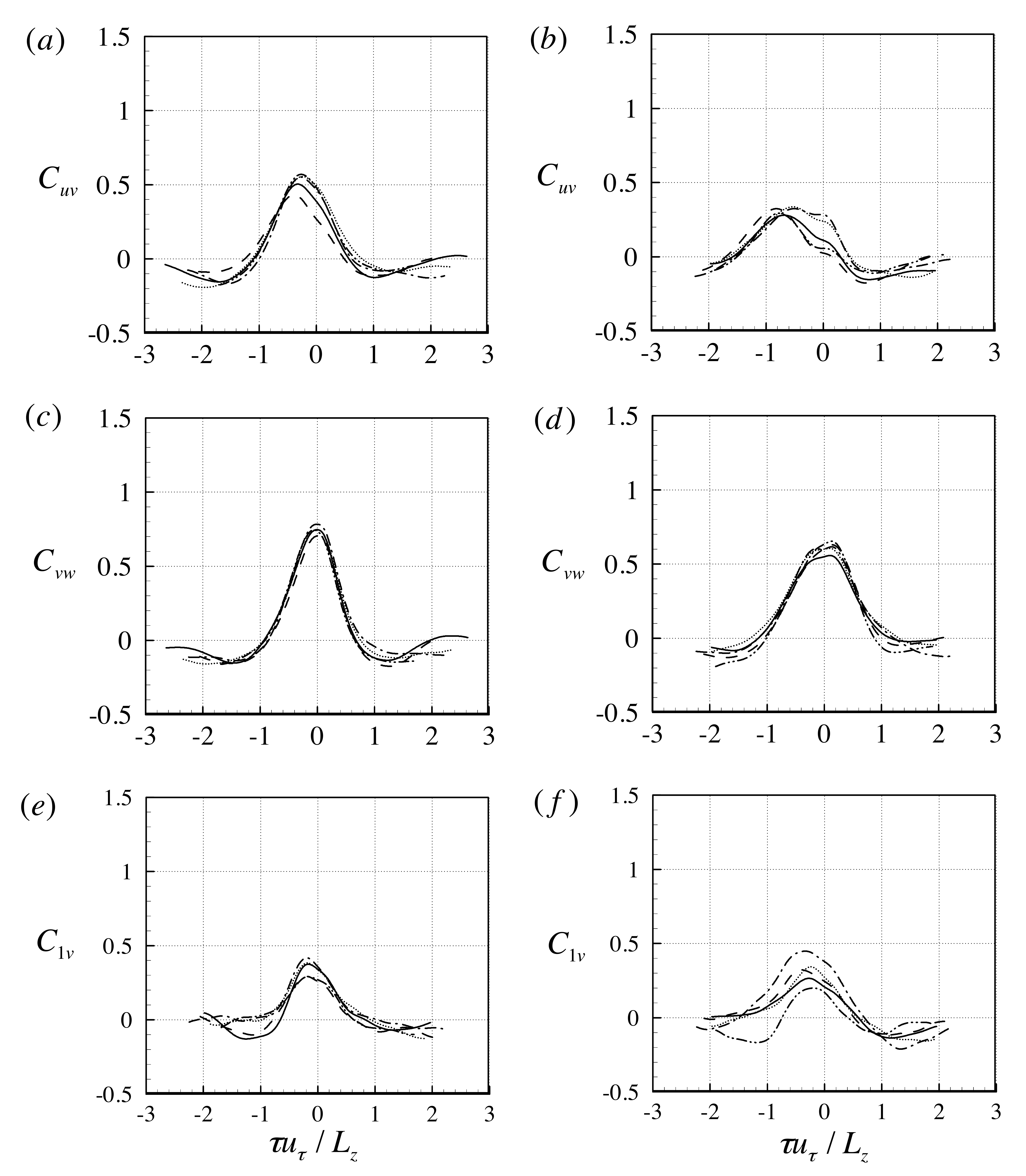}
\caption{Cross-correlation functions: $(a,b)$ $C_{uv}(\tau)$; $(c,d)$ $C_{vw}(\tau)$; $(e,f)$ $C_{1v}(\tau)$. In $(a,c,e)$, \protect \solid $L950a$; \protect \dashed $L950b$; \protect \dashdot $L1800a$; \protect \dashdotdot $L1800b$; \protect \dotted $L1800c$, while, in $(b,d,f)$, \protect \solid $SL950a$; \protect \dashed $SL950b$; \protect \dashdot $SL1800a$; \protect \dashdotdot $SL1800b$; \protect \dotted $SL1800c$.}\label{fig8}
\end{figure}

Figure \ref{fig7} shows auto-correlations obtained from all the simulations concerning the energy-containing motions in the logarithmic region. All the correlation functions from both full (figures \ref{fig7}$a$,$c$,$e$) and over-damped simulations  (figures \ref{fig7}$b$,$d$,$f$) scale fairly well with the spanwise size of the attached eddies $L_z(=\lambda_{z,0})$, suggesting that the attached eddies in the minimal unit are `dynamically' self-similar with respect to the spanwise size. The auto-correlation functions $C_{uu}$ and $C_{vv}$ in figures \ref{fig7} $(a-d)$ reach zero at $\tau u_\tau/L_z \simeq \pm 1$, indicating that the single period of self-sustaining process would be roughly given by $T u_\tau/L_z\simeq 2$. As in the case of the outer attached eddies, $C_{11}$, which characterises the time scale of the meandering motion of the streak, is found to have much shorter time scale $T u_\tau/L_z\simeq 1$ than that of $C_{uu}$ and $C_{vv}$.

Figure \ref{fig8} shows several cross-correlations obtained from the same simulations. The cross-correlation $C_{uv}$ of both full (figure \ref{fig8}$a$) and over-damped simulations (figure \ref{fig8}$b$) exhibits a phase difference between $E_u$ and $E_v$ (or $E_w$), as that of the simulations concerning the outer-scaling motions (figures \ref{fig5}$c$ and $d$). A strong correlation between $E_v$ and $E_w$ is also observed in the motions in the logarithmic region, similarly to those in the outer region (figures \ref{fig8}$c$ and $d$). Finally, the cross-correlation $C_{1v}$ also confirms that amplification of $E_v$ appears a little after $E_1$ is amplified, indicating that the vortical structures in the logarithmic region are regenerated after the meandering motion (figures \ref{fig8}$e$ and $f$).

\section{Discussion}\label{sec:4}

Thus far, we have explored the dynamical behaviour of the attached eddies in the logarithmic and outer regions. The attached eddies in the minimal unit exhibits relatively large-scale temporal oscillations, `bursting' as reported by \cite{Flores2010}. The bursting is also observed in the over-damped simulations, in which the attached eddies only at the given spanwise length scale survive through their self-sustaining mechanisms. The qualitatively good comparison of instantaneous flow fields and correlation functions between the full and the over-damped simulations suggests that the bursting is presumably the reflection of the self-sustaining process of the attached eddies with the characteristic turn-over time scale given by:
\begin{equation}\label{eq:4.1}
\frac{T u_\tau}{\lambda_z}\simeq 2,
\end{equation}
where $\lambda_z$ is the spanwise size of the given attached eddies, which corresponds to the spanwise domain size of the minimal unit $L_{z}$. Here, we note that if the spanwise length scale is chosen as $\lambda_z^+\simeq 100$, (\ref{eq:4.1}) gives $T^+\simeq 200$, the well-known bursting period of the near-wall coherent structures in the minimal unit \cite[]{Hamilton1995,Jimenez2005}. On the other hand, if $\lambda_z=1.5h$ is considered, (\ref{eq:4.1}) yields $T u_\tau/h\simeq 3$, which roughly corresponds to the bursting time scale of the largest attached eddies composed of VLSMs and LSMs (see $\S$\ref{sec:3.1}). The attached eddies in-between bursts in a self-similar manner with the time scale given by (\ref{eq:4.1}) (see $\S$\ref{sec:3.2}). We note that if $\lambda_z=3y$ is chosen, the time scale given in (\ref{eq:4.1}) becomes consistent with the bursting time scale of ${T u_\tau}/{y}\simeq 6$ reported by \cite{Flores2010} who extracted this time scale from a direct numerical simulation, and, interestingly, $\lambda_z=3y$ corresponds well to the scaling $y \simeq 0.3\sim 0.4\lambda_z$, below which the computed statistics of the isolated attached eddies are found to be self-similar (see also figure \ref{fig2}).

In the present study, it is also shown that the self-sustaining process of the attached eddies involves its two dynamically interconnected structural elements: one is the long streaky motion extending over the entire streamwise domain of the minimal unit (figures \ref{fig6}$c$ and $d$), and the other is the vortical structure which would be statistically in the form of quasi-streamwise vortices given the strong correlation between the wall-normal and the spanwise velocities (figures \ref{fig5}$c$ and $d$, and \ref{fig8}$c$ and $d$). The interactive dynamics between the two elements appears to be the backbone of the self-sustaining process of the attached eddies in the logarithmic and the outer regions, and is remarkably similar to that in the near-wall region \cite[]{Hamilton1995,Schoppa2002}. It should be pointed out, however, that this should not be very surprising, given the fact that the near-wall motion in the form of the near-wall streak and the quasi-streamwise vortices would simply be the smallest attached eddy \cite[]{Hwang2015}.

\begin{figure} \vspace*{2mm}
\centering
\includegraphics[width=0.80\textwidth]{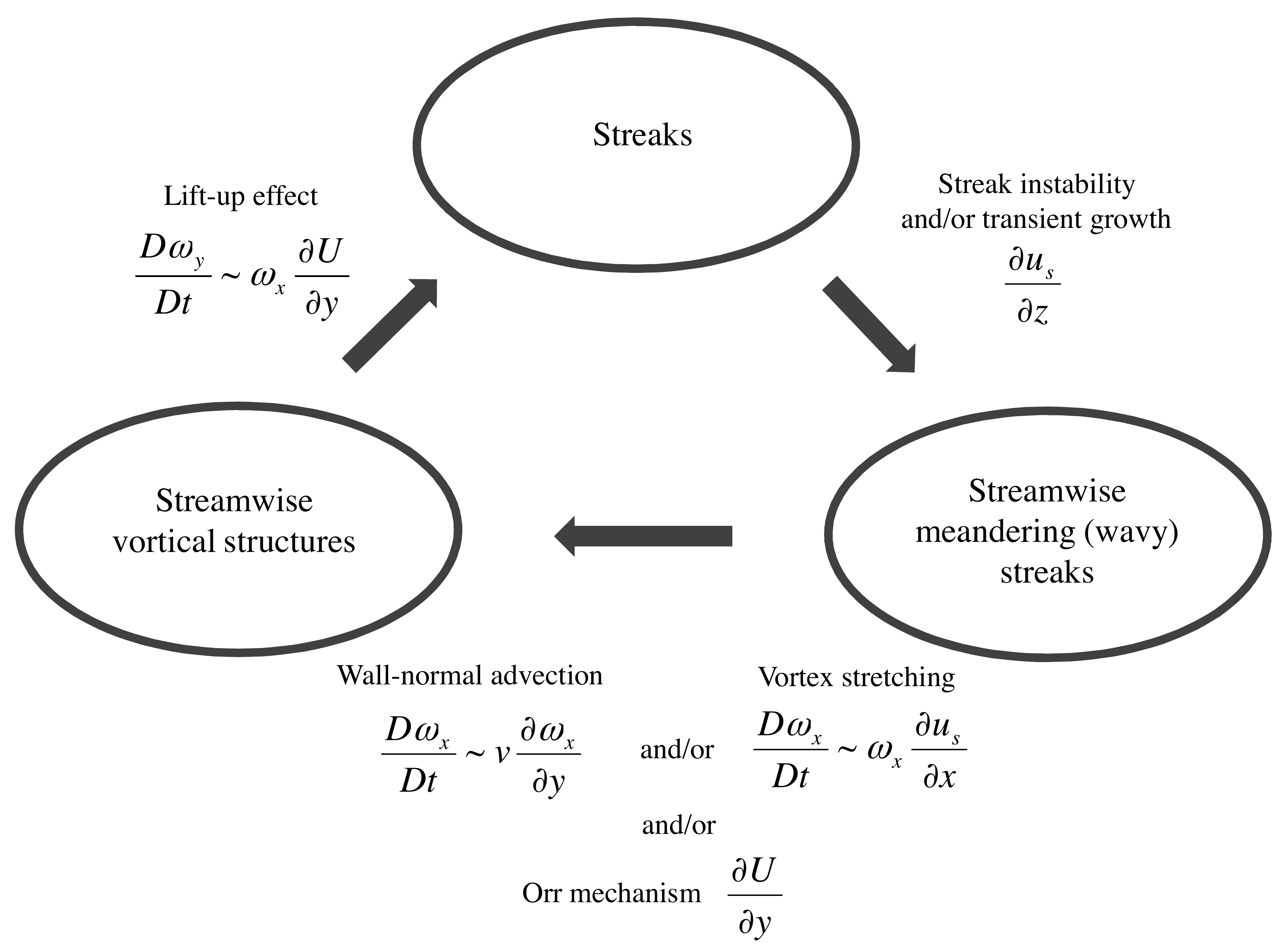}
\caption{A schematic diagram of self-sustaining process of the attached eddies \cite[originally proposed by][for the near-wall motions]{Hamilton1995}. The diagram incorporates the streak transient growth proposed by \cite{Schoppa2002} for the near-wall motions as well as the Orr mechanism proposed by \cite{Jimenez2013b,Jimenez2015}. Here, $u_s$ denotes the streamwise velocity given with the amplified streaks.}\label{fig9}
\end{figure}

Figure \ref{fig9} is a schematic diagram of the self-sustaining process of the attached eddies at a given spanwise length scale proposed in the present study based on the observations made in the previous section. This schematic diagram is originally from \cite{Hamilton1995}, and, here, we further incorporate several previous contributions in this context \cite[]{Schoppa2002,Jimenez2013b,Jimenez2015}. The streaks are amplified via the `lift-up' effect by which the streamwise vortical structures transfer energy of the mean shear to streaks (figures \ref{fig3}$c,d$ and \ref{fig4}$c,d$). The amplified streaks then undergo `rapid' meandering or oscillatory motion in the streamwise direction via streak instability or transient growth \cite[see also $\S$\ref{sec:3.2};][]{Hamilton1995,Schoppa2002,Park2011}. The quasi-streamwise vortical structures are finally regenerated with breakdown of the streak via linear and/or nonlinear mechanisms \cite[]{Hamilton1995,Schoppa2002,Jimenez2013b,Jimenez2015}. In order to provide more solid evidence on the existence of these physical processes, we now introduce two additional numerical experiments, designed to examine the lift-up effect (the left branch from streamwise vortical structures to streaks in figure \ref{fig9}) and the streak meandering and regeneration of the streamwise vortical structures (the right and lower branches from streaks to streamwise vortical structures in figure \ref{fig9}), respectively.

\subsection{Streak amplification: lift-up effect}\label{sec:4.1}

The `lift-up' effect has been very well understood as the robust mechanism of generation of the streaky motions both in transitional \citep{Ellingsen75,Landahl1980,Gustavsson1991,Butler1992,Reddy1993,Schmid01} and turbulent flows \citep{Landahl1990,Butler1993,Kim2000,Chernyshenko2005,delAlamo2006,Cossu2009,Pujals2009,Hwang2010,Hwang2010b,Willis2010}. From a vortex dynamical viewpoint, it simply represents tilting of the streamwise vortices by mean shear: i.e.
\begin{equation}\label{eq:4.2}
\frac{D\omega_y}{Dt} \sim \frac{dU}{dy} \omega_x,
\end{equation}
where $D/Dt$ is the material time derivative, $\omega_x$ and $\omega_y$ are the streamwise and wall-normal vorticities, respectively. The `lift-up' effect is an important origin of the `non-normality' of the linearised Navier-Stokes operator, leading to a large amplification of an initial condition as well as a body forcing containing a significant amount of the wall-normal velocity component. Therefore, the non-modal stability analysis \citep{Schmid01,Schmid2007}, which quantifies the amplification mechanism of the stable linearised Navier-Stokes operator, has been a popular methodology of most of the previous investigations. However, direct relevance of the `lift-up' effect to fully-developed turbulence has been very rarely shown, except by \cite{Kim2000} who demonstrated its importance on the near-wall turbulence at a low Reynolds number by performing a direct numerical simulation without the off-diagonal term of the Orr-Sommerfeld-Squire operator representing (\ref{eq:4.2}).

The goal of this section is to demonstrate that the `lift-up' effect is the essential part of the self-sustaining process of the attached eddies in the logarithmic and outer regions with a similar approach of \cite{Kim2000}. To this end, we modify their approach to suppress the lift-up effect only at a prescribed spanwise length scale. In the minimal unit, the streak amplification by the lift-up effect is found to be dominant at infinitely long streamwise wavelength (figures \ref{fig6}$c$ and $d$). To artificially suppress this lift-up effect at zero streamwise wavenumber ($k_x=0$), the following modified streamwise momentum equation is solved with the minimal unit:
\begin{eqnarray}\label{eq:4.3}
&&\frac{\partial \bar{u}}{\partial t}+(\mathbf{u} \cdot \nabla)\bar{u}-\Big([\hat{v}e^{ik_z z}+\hat{v}^*e^{-ik_z z}]\Big|_{(k_x,k_z)=(0,\frac{2\pi}{L_z})}\Big)\frac{d\overline{U}}{dy} \nonumber \\
&&=-\frac{1}{\rho}\frac{\partial p}{\partial x}+(\nu+\nu_T)\nabla^2 \bar{u},
\end{eqnarray}
where the superscript $^*$ denotes the complex conjugate, $\bar{u}$ is the streamwise velocity, and $\overline{U}$ is the streamwise velocity averaged along the streamwise and spanwise directions at each time. In (\ref{eq:4.3}), the energy extraction mechanism from the mean shear $\overline{U}$ by the wall-normal velocity (i.e. lift-up effect) is now artificially eliminated for the streamwise uniform streaky motions ($k_x=0$) at $\lambda_z=L_z$ by adding the third term in the left-hand side. We also note that this is equivalent to removing the off-diagonal term of the Orr-Sommerfeld-Squire operator for $k_x=0$ and $k_z=\pm 2\pi/L_z$, as in the approach by \cite{Kim2000} whose numerical solver is written in the wall-normal velocity and vorticity form.

\begin{figure} \vspace*{2mm}
\centering
\includegraphics[width=0.78\textwidth]{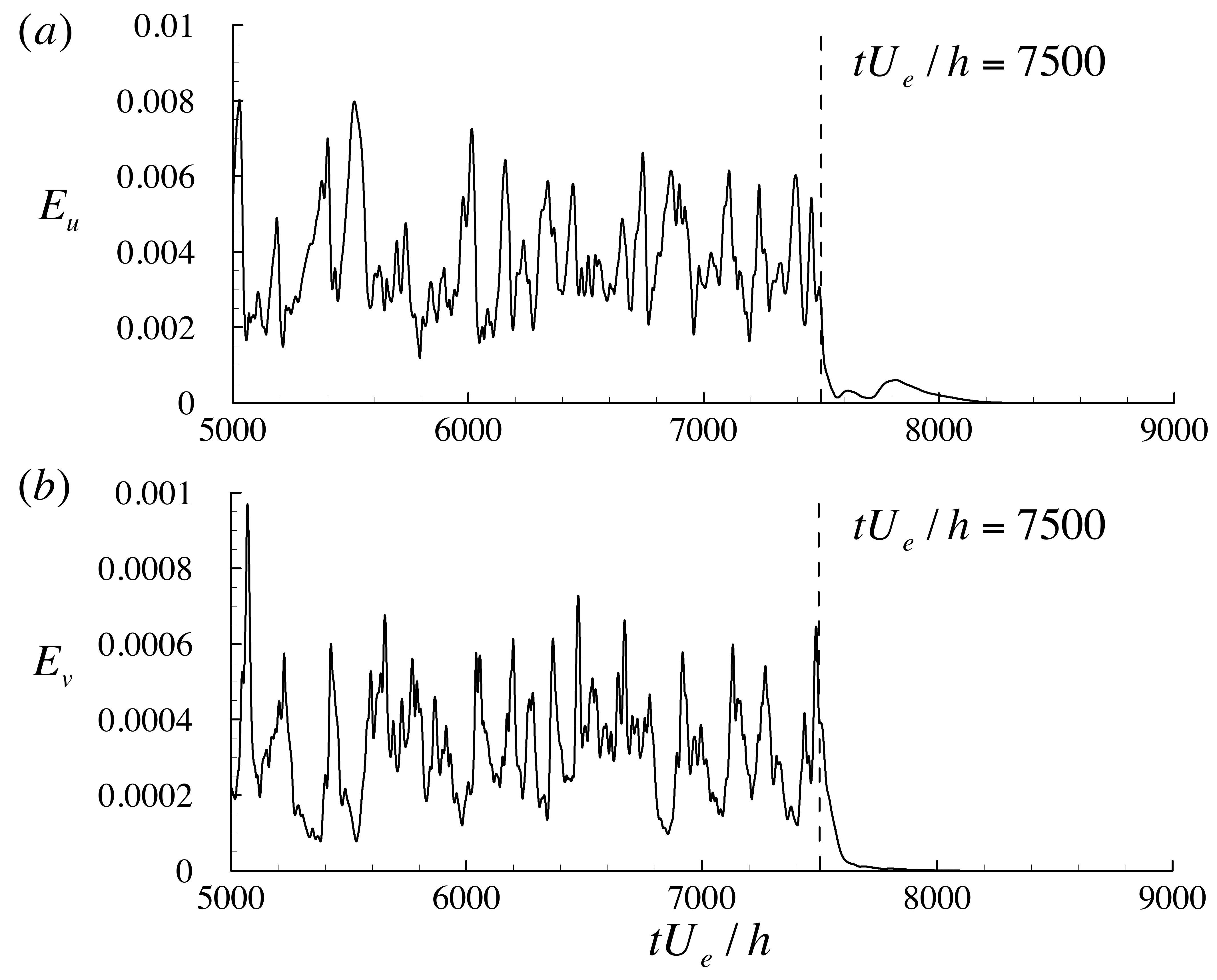}
\caption{Time trace of $E_u$ and $E_v$ of $SO950$. At $tU_e/h=7500$ ($U_e$ is the centreline velocity of the laminar velocity profile with the same volume flux), the streamwise momentum equation is replaced by (\ref{eq:4.3}). Here, $E_u$ and $E_v$ are normalised with $U_e$.}\label{fig10}
\end{figure}

The designed numerical experiment is first performed with $SO950$ in which only the largest attached eddies sustain themselves without any smaller-scale turbulent motions. Figure \ref{fig10} shows time trace of $E_u$ and $E_v$. Initially, a normal $SO950$ simulation is performed, yielding the non-trivial fluctuation of $E_u$ and $E_v$ via the self-sustaining process. At $tU_e/h=7500$ ($U_e$ is the centreline velocity of the laminar velocity profile with the same volume flux), {the streamwise momentum equation} is replaced by (\ref{eq:4.3}). Immediately after this time, $E_u$, which characterises the temporal evolution of the streak (the VLSM in this case), very rapidly decays, indicating that the streaky motion is very quickly destroyed by implementing (\ref{eq:4.3}) (figure \ref{fig10}$a$). We note that, in $tU_e/h \in [7500,7600]$, this decay of $E_u$ appears to be even faster than $E_v$, finally resulting in complete suppression of the self-sustaining process in $SO950$. {The very rapid destruction of the streaky motion by implementing (\ref{eq:4.3}) is consistently observed with a few other initial conditions.} This clearly suggests that the amplification of the long streaky motion is indeed governed by the `lift-up' effect, consistent with a number of previous theoretical predictions along this line \cite[]{delAlamo2006,Pujals2009,Hwang2010b}.

\begin{figure} \vspace*{2mm}
\centering
\includegraphics[width=0.845\textwidth]{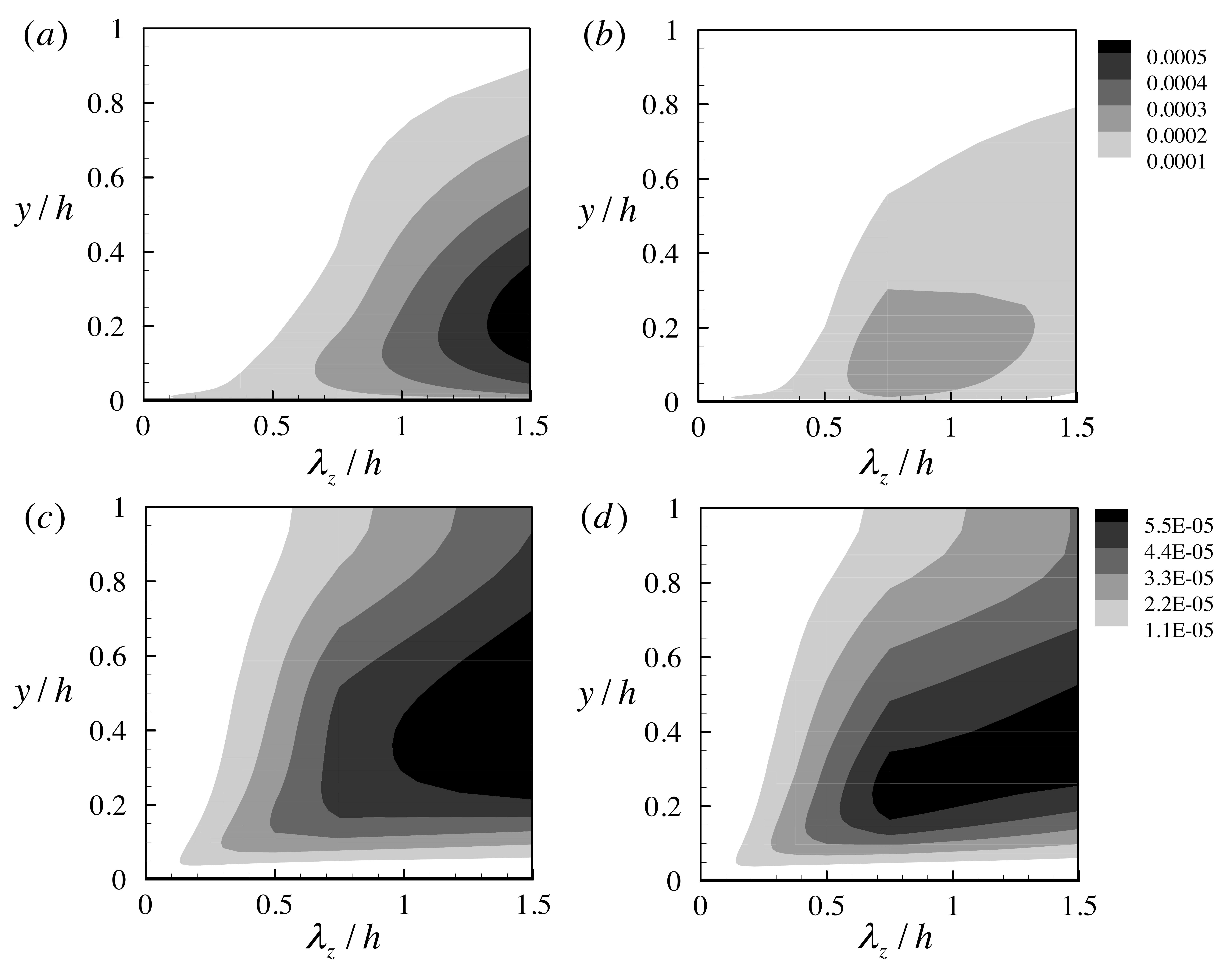}
\caption{One-dimensional spanwise wavenumber spectra of $(a,b)$ the streamwise and $(c,d)$ the wall-normal velocities: $(a,c)$ $O950$; $(b,d)$ $O950$ with (\ref{eq:4.3}). Here, the spectra are normalised with $U_e$ for the purpose of comparison.}\label{fig11}
\end{figure}

\begin{figure} \vspace*{2mm}
\centering
\includegraphics[width=0.85\textwidth]{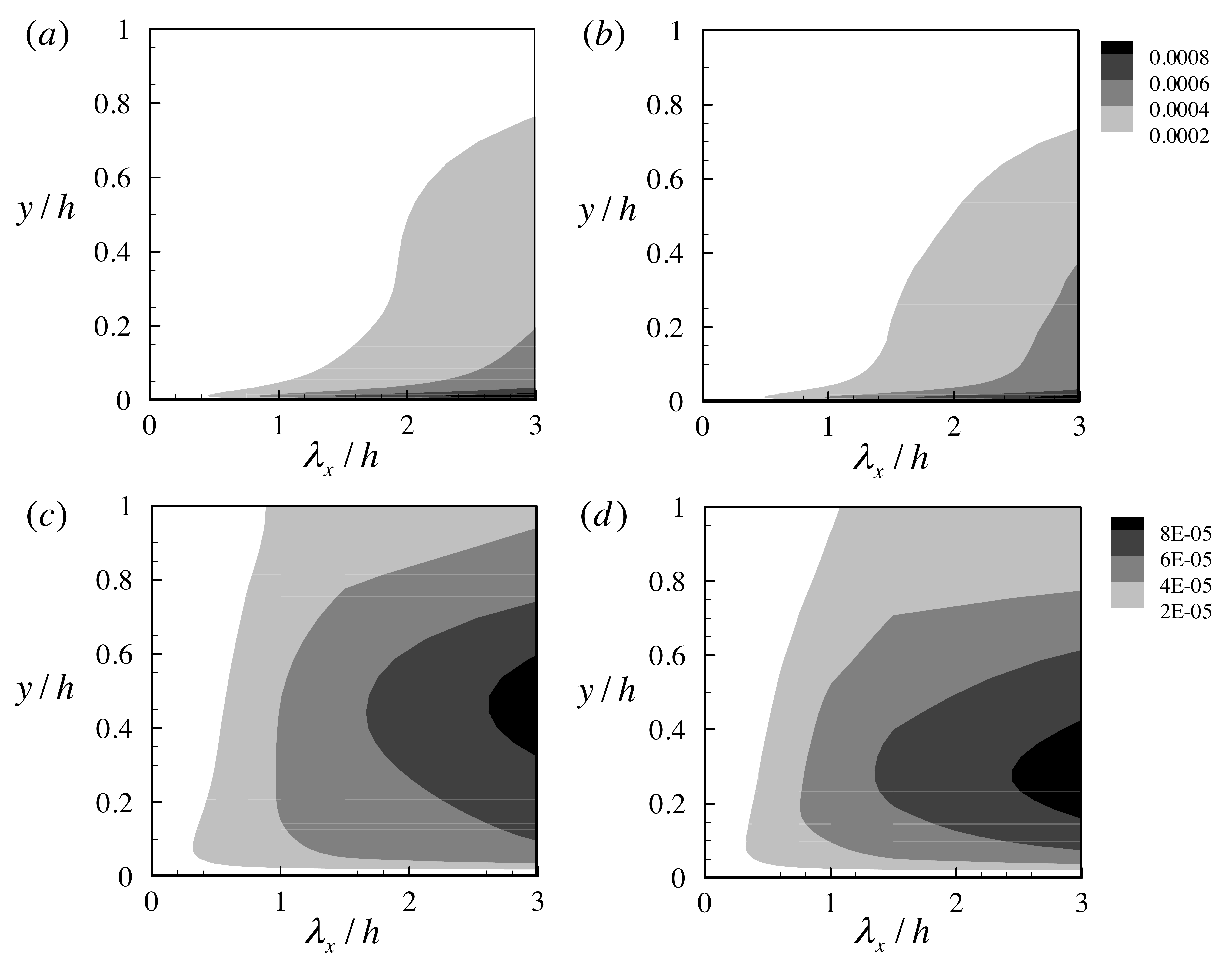}
\caption{One-dimensional streamwise wavenumber spectra of $(a,b)$ the streamwise and $(c,d)$ the wall-normal velocities: $(a,c)$ $O950$; $(b,d)$ $O950$ with (\ref{eq:4.3}). Here, the spectra are normalised with $U_e$ for the purpose of comparison.}\label{fig12}
\end{figure}

The numerical experimentation with (\ref{eq:4.3}) is also applied to $O950$ in which the largest attached eddies are resolved with turbulent motions at other length scales. Not surprisingly, $O950$ is not completely suppressed by (\ref{eq:4.3}), as the self-sustaining mechanisms at other scales would still be undisturbed. In figure \ref{fig11}, we compare the spanwise wavenumber spectra of $O950$ (figures \ref{fig11}$a$ and $c$) with those of $O950$ with (\ref{eq:4.3}) (figures \ref{fig11}$b$ and $d$). The streamwise velocity spectra of $O950$ with (\ref{eq:4.3}) clearly shows a significant amount of reduction of the streamwise turbulent kinetic energy at $\lambda_z=1.5h$ compared to those of $O950$ (figures \ref{fig11}$a$ and $b$), indicating that the streaky motion at $\lambda_z=1.5h$ (VLSM) under full turbulence is also destroyed by inhibiting the lift-up effect. This also appears to yield reduction of the wall-normal turbulent kinetic energy at $\lambda_z=1.5h$ for $y>0.5h$ (figures \ref{fig11}$c$ and $d$), as this length scale corresponds to that of the streamwise vortical structures involved in the self-sustaining process at $\lambda_z=1.5h$ (see also (\ref{eq:1.1b}) with $\lambda_z=1.5h$). However, it is quite interesting to note that the wall-normal turbulent kinetic energy at other spanwise wavelengths, particularly the ones near $\lambda_z\simeq 0.75h$, appears to be significantly amplified with the removal of the streaky motions at $\lambda_z=1.5h$, and generates additional energy even at $\lambda_z=1.5h$. This suggests that the non-trivial scale interactions exist between the attached eddies at neighbouring length scales, although the detailed investigation of this issue is beyond the scope of the present study. Inspection of the streamwise wavenumber spectra shown in figure \ref{fig12} also allows us to reach the same conclusion. Since (\ref{eq:4.3}) only applies to the streamwise uniform streaky motion (i.e. $k_x=0$), the spectra of the streamwise velocity show only a little change (figures \ref{fig11}$a$ and $b$). However, with the {destruction} of the streaky motions uniform along the streamwise direction, the spectra of the wall-normal velocity at $\lambda_x \simeq 3h$ shows a considerable amount of reduction of energy for $y>0.5h$ (figures \ref{fig11}$c$ and $d$), indicating that the streamwise vortical structures associated with the self-sustaining process at $\lambda_z=1.5h$ are destroyed. We note that the increased energy at $\lambda_x \simeq 1.5h$ in the streamwise wavenumber spectra of the wall-normal velocity should be associated with the increased energy of the wall-normal velocity near $\lambda_z\simeq 0.75h$ (figures \ref{fig11}$d$).

Finally, it has been robustly found that suppressing the lift-up effect through (\ref{eq:4.3}) yields a considerable amount of turbulent skin-friction reduction, as reported in table \ref{tab2} showing that the smaller computational box is, the larger drag reduction is achieved. Considerable amounts of turbulent skin-friction reduction by suppressing the lift-up effect here suggests that generation of turbulent skin-friction at high Reynolds numbers would be associated with the momentum transfer to the wall by the self-sustaining processes of the attached eddies in the logarithmic and outer region. This result also indicates that applying a linear flow control, e.g. with the approaches in \cite{Kim2007}, to relatively large-scale attached eddies would be a promising strategy for turbulent drag reduction. Despite many interesting further issues, the detailed investigation on the relation between turbulent skin friction and self-sustaining process (or lift-up effect) in the logarithmic and outer regions is beyond the scope of the present study, and is currently a subject of on-going investigation.

\subsection{Streak breakdown and regeneration of streamwise vortical structures}\label{sec:4.2}

\begin{table}
  \begin{center}
\def~{\hphantom{0}}
  \begin{tabular}{lclc}
   Case    & $~\Delta C_f~(\%)~$ & $~~~~$Case    & $~\Delta C_f~(\%)~$  \\[3pt]

   \hline
   $L950a~$ & $~20\%~$ & $~~~~L1800a~$ & $~30\%~$\\
   $L950b~$ & $~13\%~$ & $~~~~L1800b~$ & $~25\%~$\\
   $O950~$  & $~8\%~$ & $~~~~L1800c~$ & $~18\%~$\\ [3pt]
  \end{tabular}
  \caption{Skin-friction reduction $\Delta C_f$ by suppressing the lift-up effect with (\ref{eq:4.3}).}
  \label{tab2}
  \end{center}
\end{table}

In contrast to the `lift-up' effect which has been {an} issue of a number of previous studies, the breakdown of amplified streaks and the subsequent regeneration of the streamwise vortical structures (see figure \ref{fig9}) have been much less studied, even for the near-wall motions. In the case of the near-wall motions, which would be the smallest attached eddies \cite[]{Hwang2015}, the amplified streak typically experiences a rapid sinuous meandering or oscillatory motion before the streamwise vortices are generated \cite[]{Hamilton1995,Schoppa2002}. This sinuous meandering motion is a consequence of the streak instability or transient growth \cite[]{Hamilton1995,Schoppa2002}, although, in practice, distinguishing one from another appears to be almost impossible, given the fact that both of the processes is basically a consequence of the interaction with the same sinuous-mode instability \cite[]{Hoepffner2005}. It should also be mentioned that the sinuous mode instability essentially originates from the `spanwise shear' generated by the amplified streak \cite[i.e. $\partial u_s /\partial z$ where $u_s$ is the streamwise velocity with the amplified streak; see also][]{Park95,Cossu2002,Hoepffner2005,Park2011}. It is therefore physically unmeaningful to analyse this process using the approaches based on the linearsed Navier-Stokes equation around `mean shear', which does not carry any relevant physics on this mechanism.

The sinuous meandering motion of the amplified streak feeds a small amount of the streamwise vortical structures, and they are subsequently amplified via nonlinear mechanisms \cite[]{Hamilton1995,Schoppa2002}. \cite{Hamilton1995} proposed that wall-normal advection of the streamwise vortices would be the leading nonlinear regeneration mechanism based on a minimal Couette flow simulation ($D\omega_x/Dt \sim v \partial \omega_x/\partial y$; figure \ref{fig9}). On the other hand, with a minimal channel flow simulation, \cite{Schoppa2002} showed that stretching of the streamwise vortices by the streamwise meandering (or wavy) motion caused by the streak instability or transient growth would play an important role in this process ($D\omega_x/Dt \sim \omega_x \partial u/\partial x$; figure \ref{fig9}). More recently, \cite{Jimenez2013b,Jimenez2015} proposed that the Orr mechanism, a process by which the wall-normal velocity takes energy from the mean shear, also appears to play a role in this process especially for the motions in the logarithmic and outer regions. At this moment, it is quite difficult to convincingly argue which of the mechanisms would be dominant or most important among the three, especially for the generation of the streamwise vortical structures in the logarithmic and outer regions. However, it does not appear that solely the Orr mechanism \cite[]{Jimenez2013b,Jimenez2015}, which merely concerns the amplification of the wall-normal velocity only, would play a dominant role, given the strong correlation between the wall-normal and spanwise velocities (figures \ref{fig5}$c,d$ and \ref{fig8}$c,d$).

In the present study, it has been found that the rapid sinuous meandering motion of the amplified streak and the subsequent generation of the streamwise vortical structures are also the robust features of the self-sustaining processes in the logarithmic and outer regions (figures \ref{fig3}$c$, \ref{fig4}$c$, \ref{fig6}$c,d$, and \ref{fig8}$e,f$). This suggests that the streak instability, which was previously analysed by \cite{Park2011} for the outer attached eddies using the Floquet theory, or transient growth would also be an important process associated with the subsequent generation of the streamwise vortical structures. It is worth mentioning that this process is intricately tangled with the subsequent nonlinear amplification of the streamwise vortical structures especially via the vortex stretching mechanism proposed by \cite{Schoppa2002}, as the vortical structures take the energy required for the amplification from the meandering (or wavy) streaky motion (i.e. $\partial u_s/\partial x$).

Given this observation, it is very tempting to postulate that the streak meandering motion, presumably caused by the streak instability or transient growth, would play a crucial role in regeneration of the streamwise vortical structures. The goal of this section is to examine this mechanism. For this purpose, here, we introduce a numerical experiment which artificially damps the sinuous meandering motion of the amplified streak. Since the streak meandering motion is dominant at $\lambda_x\simeq L_x$ in the minimal unit (figures \ref{fig3}$c$ and \ref{fig4}$d$), we implement a damping of this wave component by modifying the right-hand side of the discretised streamwise momentum equation at each Runge-Kutta substep, {such that:
\begin{equation}\label{eq:4.4}
\widehat{\mathrm{RHS}}_{x} (y;k_x=\frac{2\pi}{L_x},k_z=\pm\frac{2\pi}{L_z})~\rightarrow~\mu \widehat{\mathrm{RHS}}_{x} (y;k_x=\frac{2\pi}{L_x},k_z=\pm\frac{2\pi}{L_z}),
\end{equation}}
where $\mu$ is the factor, which should be $\mu<1$ for damping. Here, we again stress that this technique is implemented only to the streamwise momentum equation to only suppress the meandering motion of $u$.

\begin{figure} \vspace*{2mm}
\centering
\includegraphics[width=0.78\textwidth]{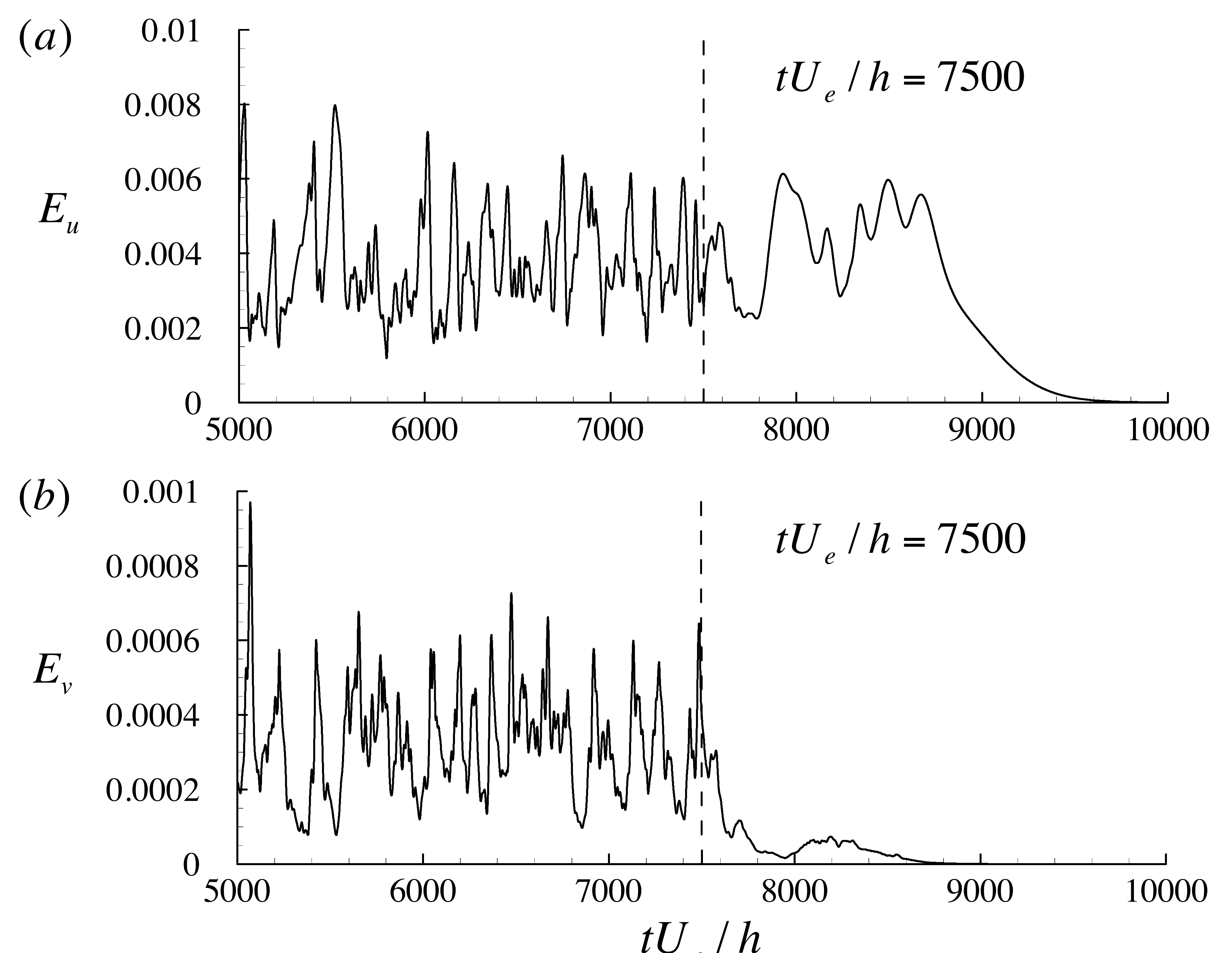}
\caption{Time trace of $E_u$ and $E_v$ of $SO950$. At $tU_e/h=7500$, (\ref{eq:4.4}) is applied. Here, $E_u$ and $E_v$ are normalised with $U_e$.}\label{fig13}
\end{figure}

The damping technique (\ref{eq:4.4}) is applied to both $O950$ and $SO950$ by considering a few different $\mu(=0.85,0.90,0.95)$. However, the results are found to {be} qualitatively independent of the choice of $\mu$, thus the results only for $\mu=0.95$ are reported in the present study. Figure \ref{fig13} shows time trace of  $E_u$ and $E_v$ of $SO950$ in which (\ref{eq:4.4}) is applied at $tU_e/h=7500$. Immediately after (\ref{eq:4.4}) is applied, $E_v$ decays very quickly. {Examination with a few other initial conditions reveals that this behaviour is qualitatively independent of the initial condition,} suggesting that the streak meandering motion observed in $SO950$ is indeed directly involved in generation of the streamwise vortical structures.
In $SO950$, both of $E_u$ and $E_v$ eventually decay to zero after application of (\ref{eq:4.4}), but, in this case, it is quite interesting to note that $E_u$ persists for a substantially long time. This certainly differs from the behaviour of $E_u$ observed after applying (\ref{eq:4.3}) (compare figure \ref{fig13}$a$ with \ref{fig10}$a$), implying that the lift-up effect is a very strong amplification process of the streamwise velocity (or streak) even with a small amplitude of the vortical structures.

\begin{figure} \vspace*{2mm}
\centering
\includegraphics[width=0.88\textwidth]{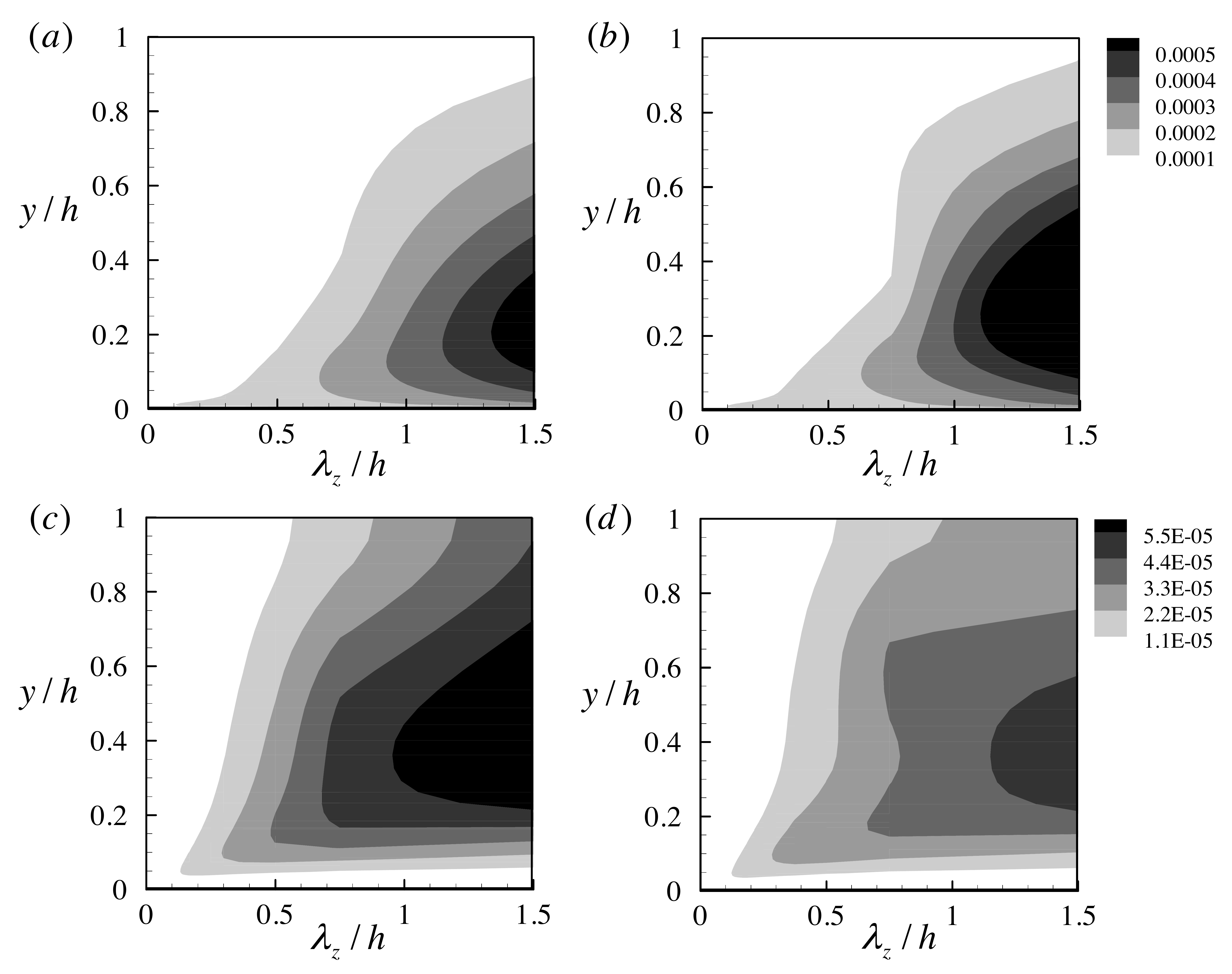}
\caption{One-dimensional spanwise wavenumber spectra of $(a,b)$ the streamwise and $(c,d)$ the wall-normal velocities: $(a,c)$ $O950$; $(b,d)$ $O950$ with (\ref{eq:4.4}). Here, the spectra are normalised with $U_e$ for the purpose of comparison.}\label{fig14}
\end{figure}

\begin{figure} \vspace*{2mm}
\centering
\includegraphics[width=0.85\textwidth]{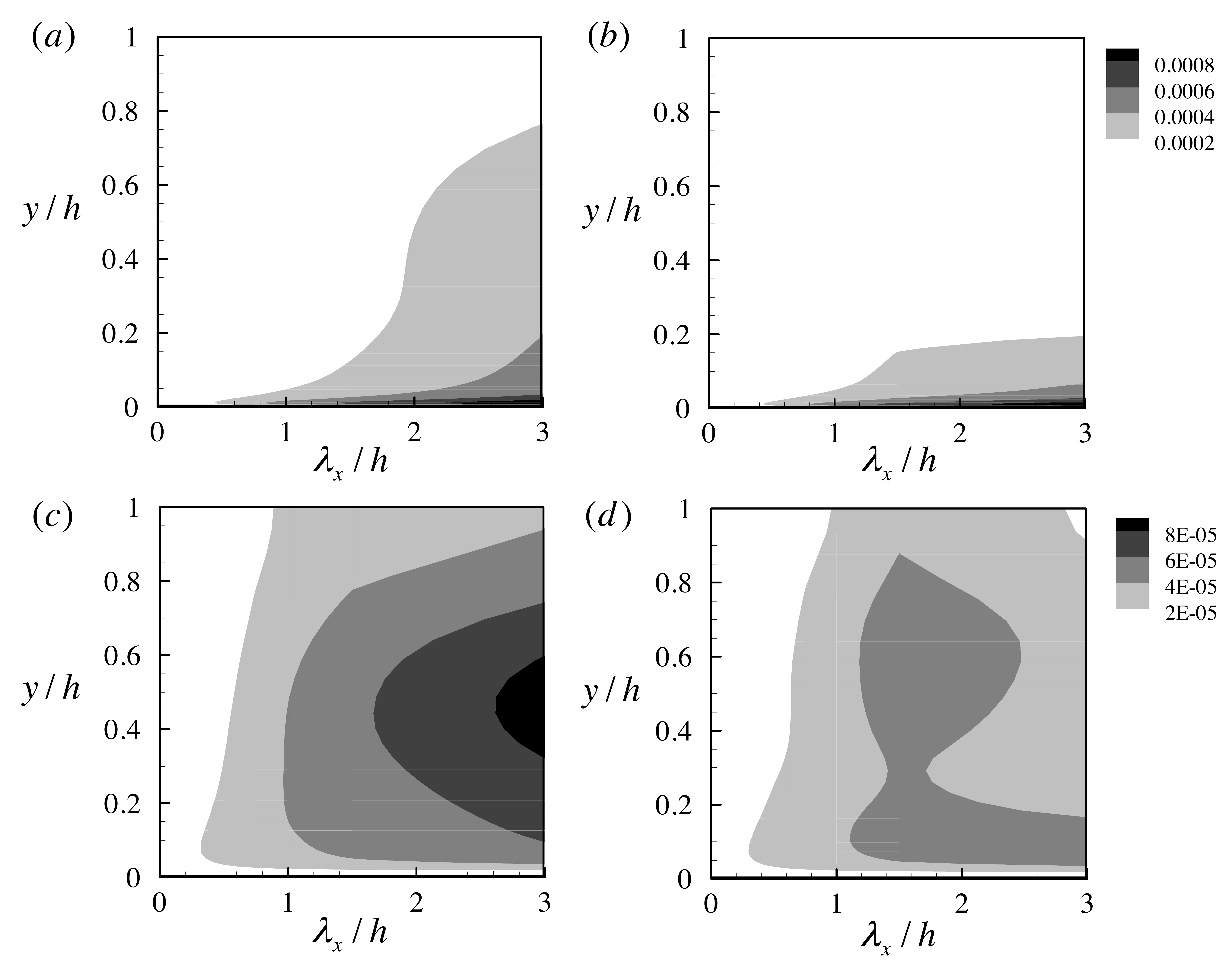}
\caption{One-dimensional streamwise wavenumber spectra of $(a,b)$ the streamwise and $(c,d)$ the wall-normal velocities: $(a,c)$ $O950$; $(b,d)$ $O950$ with (\ref{eq:4.4}). Here, the spectra are normalised with $U_e$ for the purpose of comparison.}\label{fig15}
\end{figure}

The damping technique (\ref{eq:4.4}) is also applied to $O950$. The streamwise and spanwise wavenumber spectra of the streamwise and the wall-normal velocity are compared with those of $O950$, as respectively reported in figures \ref{fig14} and \ref{fig15}.
Consistent with the case of $SO950$, application of (\ref{eq:4.4}) to $O950$ yields significant reduction of the wall-normal turbulent kinetic energy at $\lambda_z=1.5h$ (figure \ref{fig14}$c$ and $d$), despite the increase of the streamwise turbulent kinetic energy in the spanwise wavenumber spectra (figures \ref{fig14}$a$ and $b$; see also the next paragraph for a further discussion). The streamwise wavenumber spectra of the wall-normal turbulent kinetic energy also clearly exhibits a significant amount of reduction of the energy at $\lambda_x=3h$ (figure \ref{fig15}$c$ and $d$), indicating that the artificial damping of the streak meandering motion significantly destroys the generation mechanism of the streamwise vortical structures at this wavelength.

Despite the consistent result with $SO950$, in $O950$, application of (\ref{eq:4.4}) does not appear to completely destroy the structures at $\lambda_z=1.5h$. Although the streamwise wavenumber spectra of the streamwise velocity shows a large amount of diminish of the energy at $\lambda_z=1.5h$ (figures \ref{fig15}$a$ and $b$), confirming that (\ref{eq:4.4}) indeed inhibits the meandering motion of the streak at this wavelength, the increase of the streamwise turbulent kinetic energy in the spanwise wavenumber spectra (figure \ref{fig14}$b$) implies that the streamwise uniform streaky motion (i.e. $k_x=0$ mode in the minimal unit) is amplified on average. This also appears to be consistent with little change of the skin-friction drag, despite the significant destruction of the vortical structures at $\lambda_x=3h$.

The intensification of the energy of the streamwise uniform motion is surprising at least to us, given the complete suppression of the self-sustaining process in $SO950$. {The increase of the energy of the streamwise uniform motion might be explained by the lack of a proper breakdown mechanism through the streak instability or transient growth artificially suppressed, as weaker vortices could also sustain stronger streaks which can break down at shorter streamwise wavelengths with the elevated breakdown threshold. However, the persisting streaky motion even with the damaged breakdown mechanism can also suggest that there may exist some additional feeding mechanisms of the streamwise vortical structures.} Currently, we do not gain precise understanding on these mechanisms, although they might be from the scale interaction or from the Orr-mechanism proposed by \cite{Jimenez2013b,Jimenez2015}. However, in any case, the result of the present numerical experiment directly supports the notion that the streak instability (or transient growth) and the following nonlinear vortex stretching of the streamwise vortical structures play a crucial role in the formation of the vortical structures in the logarithmic and the outer regions as well as in the determination of their streamwise length scale \cite[]{Schoppa2002,Park2011,Hwang2015}.

\section{Concluding remarks}\label{sec:5}
Thus far, we have investigated the self-sustaining process of the energy-containing motions in the logarithmic and outer regions (the attached eddies) by examining their dynamical behaviour in the minimal unit. The present study is summarised as follow:
\begin{enumerate}
\item The attached eddies at a given spanwise length scale in the minimal unit exhibit a statistically recurrent dynamical behaviour which has been called `bursting' \cite[]{Flores2010}. The bursting is also observed in the over-damped simulations in which the attached eddies survive only through their self-sustaining mechanisms, and is remarkably similar to that in the full simulations, indicating that the bursting is the reflection of the self-sustaining process of the attached eddies. For the attached eddies at the given spanwise length scale $\lambda_z$, the time period of the bursting (i.e. the self-sustaining process) is found to scale with $T u_\tau/\lambda_z \simeq 2$, suggesting that the statistically self-similar attached eddies in the logarithmic region \cite[]{Hwang2015} is also `dynamically' self-similar.

\item It is shown that the self-sustaining process of the attached eddies in the logarithmic and outer regions is very similar to that of the near-wall motions \cite[]{Hamilton1995,Schoppa2002}, which would be the smallest attached eddies \cite[]{Hwang2015}. The {attached} eddies in the minimal unit in the logarithmic and outer regions, composed of two elements, the streak and the streamwise vortical structures, sustain themselves through the interactive dynamics between the two: 1) the streak is significantly amplified by the streamwise vortical structures via the lift-up effect; 2) the amplified streak subsequently experiences very rapid meandering motion along the streamwise direction; 3) the meandering streaks break down with regeneration of new streamwise vortical structures.

\item To provide more convincing evidence on the existence of the proposed self-sustaining process in the logarithmic and outer regions, two numerical experiments are further performed. One is to artificially suppress the lift-up effect by modifying the technique introduced by \cite{Kim2000}, and the other is to artificially damp out the streak meandering motion, which would probably be a consequence of the streak instability or transient growth \cite[]{Park2011}. It is shown that the artificial suppression of the lift-up effect inhibits the amplification of the streak and subsequently suppresses the self-sustaining process of the attached eddies. Also, the artificial inhibition of the streak meandering motions destroy the generation of the streamwise vortical structures, significantly affecting the self-sustaining process. The numerical experiments also reveal that there are non-trivial scale interactions among the attached eddies.
\end{enumerate}

The importance of the present study probably lies in the identification of the self-sustaining process of the energy-containing motions in the logarithmic and outer regions, given in the form of Townsend's attached eddies. The existence of the self-sustaining process in these regions, which appears to be basically the same as that of the near-wall region, would provide some evidence on the relevance of the so-called `exact coherent structures' at high Reynolds numbers: i.e. the exact solutions of the Navier-Stokes equation in the form of unstable stationary/traveling waves {and relative periodic orbits} \cite[e.g.][]{Nagata1990,Waleffe2001,Kawahara2001,Faisst2003,Wedin2004,Jimenez2005}. At low Reynolds numbers, these solutions have been shown to form a skeleton of the basin of attraction of turbulent solution in phase space \cite[e.g.][]{Gibson2008}, and their discovery has played a crucial role in recent advancement on understanding bypass transition and low-Reynolds-number turbulence. It should be {mentioned} that the exact coherent structure, typically given in the form of a wavy streak and flanked streamwise vortices, physically represents the self-sustaining process given in figure \ref{fig9} \cite[e.g][]{Waleffe2003}. This therefore suggests that the attached eddies in the logarithmic and outer regions, which bear such a self-sustaining process, would be linked to the exact coherent structures at high Reynolds numbers {at least to some extent}.

Nevertheless, it is not yet clear how much such dynamical-system-based approaches could be extended further especially for describing the dynamics of coherent structures in high-Reynolds-number wall turbulence. Perhaps, one of the most important challenges would be description of the complex scale interactions between the self-sustaining structures at different length scales. In fact, the dynamical-system-based approaches implicitly assume that turbulence is a chaos occurring in a finite dimensional nonlinear system. Unfortunately, turbulence is, however, a chaos occurring in a spatially extended system, the dimension of which is virtually infinite, and is inherently a multi-scale phenomenon involving daunting scale interactions. Given this difficulty, the full suitability of the dynamical system approach especially to high-Reynolds-number turbulence is yet an open question, although the approach may enlighten at least some important aspects of turbulence.

\section*{Acknowledgements}
\begin{acknowledgments}
Y.H. gratefully acknowledges Prof. J. Jim\'enez for sharing his paper \cite[]{Jimenez2015} before its publication. This work was partially supported by the Engineering and Physical Science Research Council (EPSRC) in the UK (EP/N019342/1).
\end{acknowledgments}

\appendix
\section{Effect of the filtering (\ref{eq:2.1}) on time correlation functions}\label{sec:appendix}

\begin{figure} \vspace*{2mm}
\centering
\includegraphics[width=0.89\textwidth]{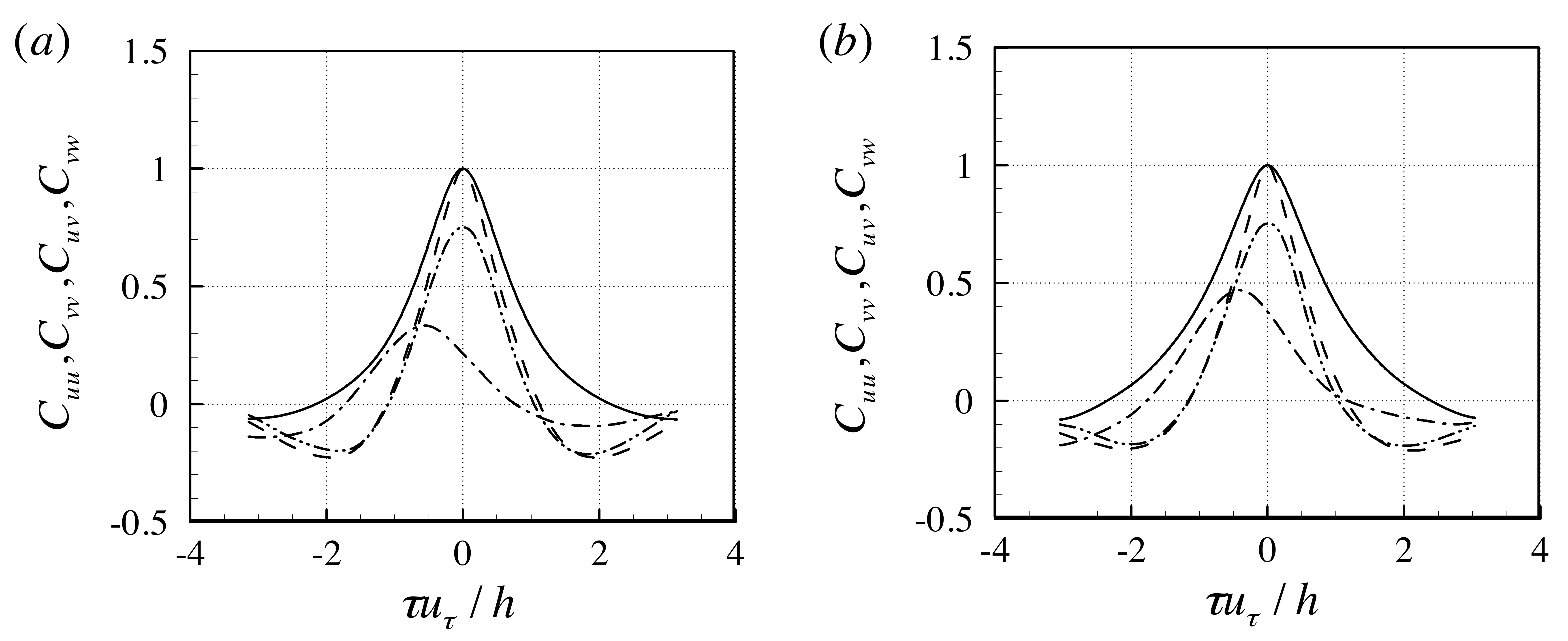}
\caption{Time correlation functions of $(a)$ $O950$ without (\ref{eq:2.1}) and $(b)$ $O950$. Here, \protect \solid, $C_{uu}(\tau)$;  \protect \dashed, $C_{vv}(\tau)$; \protect \dashdot, $C_{uv}(\tau)$; \protect \dashdotdot, $C_{vw}(\tau)$.}\label{fig16}
\end{figure}

As mentioned, all the simulations of the present study are performed with the filtering technique (\ref{eq:2.1}) which removes the two-dimensional spurious motions populating the $x$-$y$ plane in the spanwise minimal unit simulations; for further details, the reader also refers to \cite{Hwang2013} in which a detailed discussion on this spurious motion is discussed. In figure \ref{fig16}, a set of time correlation functions of $O950$ without (\ref{eq:2.1}) are compared with those of $O950$. All the presented correlation functions of $O950$ without (\ref{eq:2.1}) show good agreement with those of $O950$, except $C_{uv}(\tau)$. It appears that $C_{uv}(\tau)$ of $O950$ without (\ref{eq:2.1}) is a little larger than that of $O950$, although the peak location of $C_{uv}(\tau)$ itself shows good agreement with each other. However, this may have been expected in a way, given the nature of the filtering technique (\ref{eq:2.1}) which removes the streamwise and wall-normal motions in the $x$-$y$ plane. Overall good agreement of the correlation functions from the two simulations suggests that application of (\ref{eq:2.1}) does not significantly affect the motions in $O950$, as in our previous studies \cite[]{Hwang2013,Hwang2015}.

\section{Comparison of statistics between attached eddies in the minimal and long computational domains}\label{sec:appendixB}

\begin{figure} \vspace*{2mm}
\centering
\includegraphics[width=0.91\textwidth]{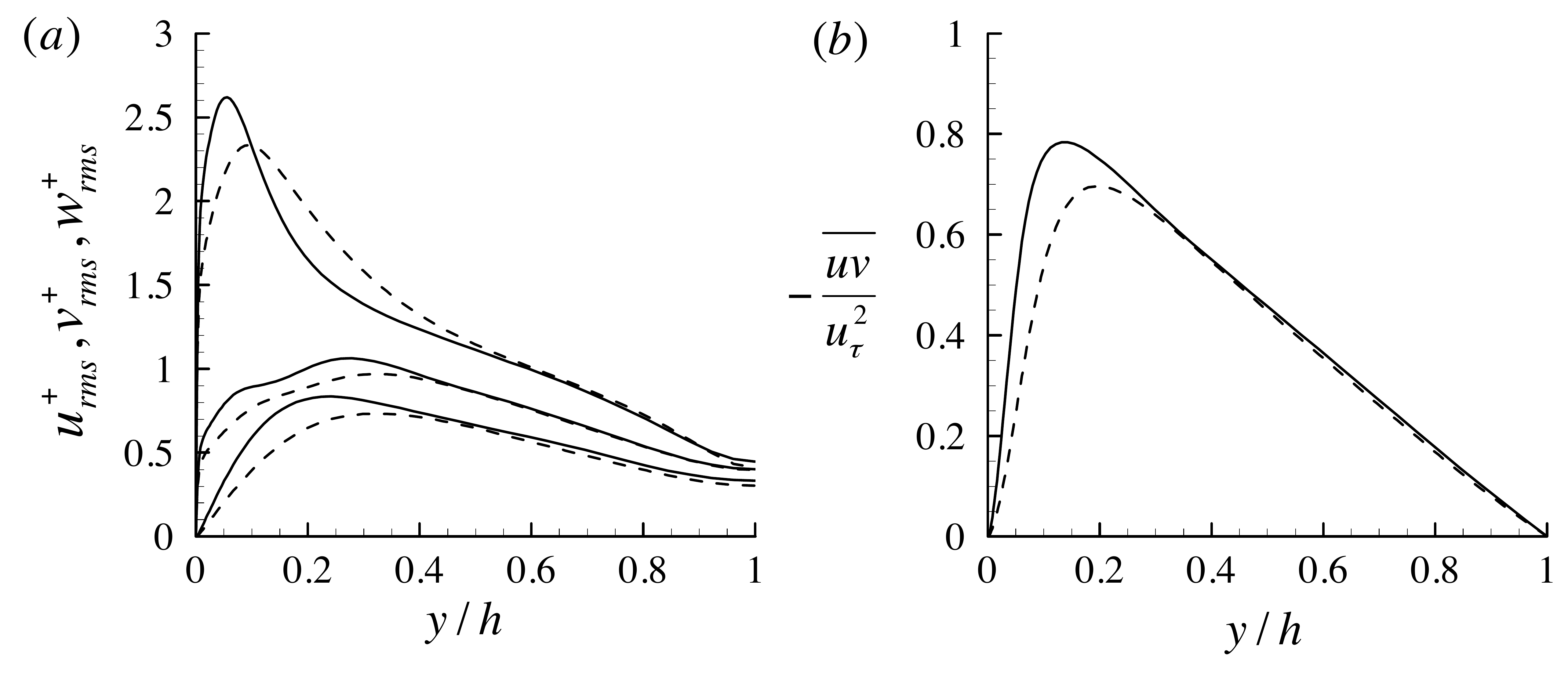}
\caption{Comparison of the second-order statistics of the self-sustaining attached eddies in the minimal unit ($SL1800c$ in the present study) with those in the long streamwise domain \cite[$L1800c$ in][]{Hwang2015}: $(a)$ the streamwise, wall-normal and spanwise velocities; $(b)$ Reynolds stress. Here, \protect \solid, $SL1800c$ in the present study; \protect \dashed, $L1800c$ in \cite{Hwang2015}.}\label{fig17}
\end{figure}

Here, we compare the statistics of the self-sustaining attached eddies in the minimal streamwise domain ($L_x=2L_z$; $SL1800c$ in the present study) with those in a long streamwise domain \cite[$L_x=8 \pi h$; $L1800c$ in][]{Hwang2015}. We note that the two simulation share the same spanwise computational domain size, $L_z=0.75h$, although the resolution of the present study is a little finer than that in \cite{Hwang2015}. Except these, all the simulation parameters of the two simulations are identical to each other. Figure \ref{fig17} shows the second-order statistics of $SL1800c$ in the present study and those of $L1800c$ in \cite{Hwang2015}. The second-order statistics obtained with the streamwise minimal domain tend to generate larger velocity fluctuations and Reynolds stress in the region relatively close to the wall, except the streamwise velocity fluctuation showing a little non-trivial difference. However, overall, the second-order statistics with the streamwise minimal domain do not considerably deviate from those with the long domain, indicating that the minimal domain does not significantly distort the second-order statistics of the interest. It should mentioned that qualitatively the same behaviour is observed in all the other minimal-unit simulations in the present study in comparison to those with sufficiently long computational domains in \cite{Hwang2015}.


\bibliographystyle{jfm}

\bibliography{yongyun}

\end{document}